\documentclass[acmtocl]{acmtrans2m}

\usepackage{enumerate}

\usepackage{amsmath}
\usepackage{amssymb}
\usepackage[mathscr]{eucal}

\usepackage{graphics}

\usepackage{xspace}

\newcommand{\IGNORE}[1]{}

\newcommand{\ie}{i.\@\,e.\@\xspace}
\newcommand{\eg}{e.\@\,g.\@\xspace}

\newcommand{\Nat}{\mathbb{N}}
\newcommand{\Int}{\mathbb{Z}}

\newcommand{\BigOh}{\operatorname{O}}

\newcommand{\set}[2]{\{#1\,:\,#2\}}

\newcommand{\remainder}{\operatorname{rem}}
\newcommand{\abs}[1]{|#1|}
\newcommand{\length}[1]{|#1|}

\newcommand{\Langle}{\mathopen{\langle\hspace{-0.2em}\langle}}
\newcommand{\Rangle}{\mathclose{\rangle\hspace{-0.2em}\rangle}}

\newcommand{\toints}[1]{\langle#1\rangle_{\Int}}
\newcommand{\tonats}[1]{\langle#1\rangle_{\Nat}}
\newcommand{\towordsints}[1]{\Langle#1\Rangle_{\Int}}
\newcommand{\towordsnats}[1]{\Langle#1\Rangle_{\Nat}}

\newcommand{\dbrackl}{\mathopen{[\hspace{-0.137em}[}}
\newcommand{\dbrackr}{\mathclose{]\hspace{-0.137em}]}}
\newcommand{\Zrepresents}[1]{\dbrackl #1\dbrackr}

\newcommand{\AND}{\mathrel{\wedge}}
\newcommand{\OR}{\mathrel{\vee}}
\newcommand{\IMPL}{\mathrel{\rightarrow}}
\newcommand{\IFF}{\mathrel{\leftrightarrow}}
\newcommand{\NEG}{\neg}

\newcommand{\divides}{{\mathrel{\vert}}}
\newcommand{\DIVIDES}{\mathbin{\big|\,}}

\newcommand{\EQ}{\mathbin{=}}
\newcommand{\NOTEQ}{\mathbin{\not=}}
\newcommand{\LESS}{\mathbin{<}}
\newcommand{\LEQ}{\mathbin{\leq}}
\newcommand{\GREATER}{\mathbin{>}}
\newcommand{\GEQ}{\mathbin{\geq}}

\newcommand{\zaeckzaeck}{{\mathrel<\joinrel\mathrel{\!\!\!>}}}
\newcommand{\ZAECKZAECK}{\mathbin{\mathrel<\joinrel\mathrel{\!\!\!>}}}

\newcommand{\CDOT}{\mathbin{\cdot}}

\newcommand{\norm}[1]{\Arrowvert#1\Arrowvert}
\newcommand{\normpos}[1]{\norm{#1}_+}
\newcommand{\normneg}[1]{\norm{#1}_-}

\newcommand{\lcm}{\operatorname{lcm}}

\newcommand{\qf}{\mathsf{QF}}
\newcommand{\pa}{\mathsf{PA}}
\newcommand{\Atoms}{{\sf A}}

\newcommand{\Terms}{{\sf T}}
\newcommand{\Divs}{{\sf D}}

\newcommand{\maxCoef}{\operatorname{max_{coef}}}
\newcommand{\maxConst}{\operatorname{max_{const}}}
\newcommand{\maxDiv}{\operatorname{max_{div}}}

\newcommand{\Prod}{\mathrm{Prod}}
\newcommand{\MULT}{\mathrm{MULT}}

\makeatletter

\def\create@abbrev#1#2#3{
  \def\c@a@def##1{
      \if ##1.
        \relax
      \else
        \@ifdefinable{\@nameuse{#1##1}}{\@namedef{#1##1}{#2##1}}
        \expandafter\c@a@def
      \fi
    }
  \c@a@def #3.
}

\create@abbrev{cal}{\mathcal}{ABCDEFGHIJKLMNOPQRSTUVWXYZ}
\create@abbrev{scr}{\mathscr}{ABCDEFGHIJKLMNOPQRSTUVWXYZ}
\create@abbrev{frak}{\mathfrak}{ABCDEFGHIJKLMNOPQRSTUVWXYZ}
\create@abbrev{rm}{\mathrm}{ABCDEFGHIJKLMNOPQRSTUVWXYZabcdefghijklmnopqrstuvwxyz
}
\create@abbrev{sf}{\mathsf}{ABCDEFGHIJKLMNOPQRSTUVWXYZabcdefghijklmnopqrstuvwxyz
}

\create@abbrev{aut}{\mathscr}{ABCDEFGHIJKLMNOPQRSTUVWXYZ}

\makeatother

\newtheorem{theorem}{Theorem}[section]

\newtheorem{corollary}[theorem]{Corollary}

\newtheorem{lemma}[theorem]{Lemma}
\newtheorem{fact}[theorem]{Fact}
\newdef{algorithm}[theorem]{Algorithm}
\newdef{definition}[theorem]{Definition}
\newdef{remark}[theorem]{Remark}
\newdef{example}[theorem]{Example}

\newcommand{\ignore}[1]{}
           
\title{Bounds on the Automata Size for Presburger Arithmetic}
            
\author{FELIX KLAEDTKE\\ETH Zurich}
            
\begin{abstract} 
  Automata provide a decision procedure for Presburger arithmetic.
  However, until now only crude lower and upper bounds were known on
  the sizes of the automata produced by this approach.  In this paper,
  we prove an upper bound on the the number of states of the minimal
  deterministic automaton for a Presburger arithmetic formula. This
  bound depends on the length of the formula and the quantifiers
  occurring in the formula.  The upper bound is established by
  comparing the automata for Presburger arithmetic formulas with the
  formulas produced by a quantifier elimination method. We also show
  that our bound is tight, even for nondeterministic automata.
  Moreover, we provide optimal automata constructions for linear
  equations and inequations.
\end{abstract}

\category{F.1.1}{Computation by Abstract Devices}{Models of
  Computation}[automata]
\category{F.4.1}{Mathematical Logic and Formal Languages}{Mathematical
  Logic}[computational logic]
            
\terms{Algorithms, Theory} 
            
\keywords{Automata-based Decision Procedures, Presburger Arithmetic,
  Quantifier Elimination, Complexity}

\begin{document}

\begin{bottomstuff} 
  This work was partially supported by the German Research Council
  (DFG) and the Swiss National Science Foundation (SNF).
  
  A preliminary version of this paper appeared at the 19th Annual IEEE
  Symposium on Logic in Computer Science (LICS'04).

  Author's address: Felix Klaedtke, ETH Zurich, Department of Computer Science,
  Haldeneggsteig 4/Weinbergstra\ss{}e, 8092 Zurich, Switzerland.
\end{bottomstuff}
            
\maketitle

\section{Introduction}
\label{INTRO}

Presburger arithmetic (PA) is the first-order theory with addition and
the ordering relation over the integers.  A number of decision
problems can be expressed in it, such as solvability of systems of
linear Diophantine equations, integer programming, and various
problems in system verification.  The decidability of PA was
established around 1930 independently by
Presburger~\citeyear{Presburger.1929,Stansifer.1984} and
Skolem~\citeyear{Skolem.1931,Skolem.1970} using the method of
quantifier elimination\@.

Due to the applicability of PA in various domains, its complexity and
the complexity of decision problems for fragments of it have been
investigated intensively.  For example, Fischer and
Rabin~~\citeyear{Fischer_Rabin.1974,Fischer_Rabin.1998} gave a double
exponential nondeterministic time lower bound on any decision
procedure for PA.  Later,~\citeN{Berman.1980} showed that the decision
problem for PA is complete in the complexity class
$\mathit{LATIME}(2^{2^{\BigOh(n)}})$, \ie, the class of problems
solvable by alternating Turing machines in time $2^{2^{\BigOh(n)}}$
with a linear number of alternations. The upper bound for PA is
established by a result from~\citeN{Ferrante_Rackoff.1979} showing
that quantified variables need only to range over a restricted finite
domain of integers.  \citeN{Graedel.1988} and~\citeN{Schoening.1997}
investigated the complexity of decision problems of fragments of PA\@.

The complexity of different decision procedures for PA has also been
studied, \eg, in
\cite{Oppen.1978,Reddy_Loveland.1978,Ferrante_Rackoff.1975,Ferrante_Rackoff.1979}\@.
For instance, \citeN{Oppen.1978} showed that Cooper's quantifier
elimination decision procedure for PA~\cite{Cooper.1972} has a triple
exponential worst case complexity in deterministic time.
\citeN{Reddy_Loveland.1978} improved Cooper's quantifier elimination
and used it for obtaining space and deterministic time upper bounds
for checking the satisfiability of PA formulas in which the number of
quantifier alternations is bounded.

Another approach for deciding PA or fragments of it that has recently
become popular is to use automata; a point that was already made
by~\citeN{Buechi.1960}\@. The idea is simple: Integers are represented
as words, \eg, using the $2$'s complement representation, and the word
automaton (WA) for a formula accepts precisely the words that
represent the integers making the formula true.  The WA can be
recursively constructed from the formula, where automata constructions
handle the logical connectives and quantifiers. 
This automata-based approach for PA led to deep theoretical insights,
\eg, the languages that are regular in any base are exactly the sets
definable in PA~\cite{Cobham.1969,Semenov.1977,Bruyere_etal.1994}\@.
More recently, the use of automata has been proposed for mechanizing
decision procedures for PA and for manipulating sets definable in
PA~\cite{Boudet_Comon.1996,Wolper_Boigelot.1995}\@.
Roughly speaking, this applied use of WAs for PA is similar to the use
of binary decision diagrams (BDDs) for propositional logic.
For example, the automata library LASH~\cite{LASH} provides tool
support for manipulating PA definable sets using automata to
represented these sets, and it has been successfully used to verify
systems with variables ranging over the integers.  Other model
checkers that use WAs for computing the potential infinite sets of
reachable states of systems with integer variables are, \eg,
FAST~\cite{FAST} and ALV~\cite{ALV}\@.

A crude complexity analysis of automata-based decision procedures for
PA leads to a non-elementary worst case complexity. Namely, for every
quantifier alternation there is a potential exponential blow-up.
However, experimental
comparisons~\cite{Ranjan_Shiple_Kukula.1998,Bartzis_Bultan.2003,Ganesh_Berezin_Dill.2002}
illustrate that automata-based decision procedures for PA often
perform well in comparison with other methods.
In~\cite{Boudet_Comon.1996}, the authors claimed that the minimal
deterministic WA for a PA formula has at most a triple exponential
number of states in the length of the formula.  Unfortunately, as
explained by~\citeN{Wolper_Boigelot.2000}, the argument used
in~\cite{Boudet_Comon.1996} to substantiate this claim is incorrect.
\citeN{Wolper_Boigelot.2000} gave an argument why there must be an
elementary upper bound on the size of the minimal deterministic WA for
a PA formula.  However, their argumentation is rather sketchy and only
indicates that there has to be an elementary upper bound.

In this paper, we rigorously prove an upper bound on the size of the
minimal deterministic WA for PA formulas and thus, answer a long open
question. Namely, for a PA formula in prenex normal form, we show that
the minimal deterministic WA has at most $2^{n^{(b+1)^{a+4}}}$ states,
where $n$ is the formula length, $a$ is the number of quantifier
alternations, and $b$ is the maximal length of the quantifier blocks.
A similar upper bound holds for arbitrary PA formulas.
This bound on the automata size for PA
contrasts with the upper bound on the automata size for the monadic
second-order logic WS1S, or even WS1S with the ordering relation
``$<$'' as a primitive but without quantification over monadic
second-order variables.  There, the number of states of the minimal WA
for a formula can be non-elementary larger than the formula's
length~\cite{Stockmeyer.1974,Reinhardt.2002}\@.
In order to establish the upper bound on the automata size for PA, we
give a detailed analysis of the deterministic WAs for formulas by
comparing the constructed WAs with the quantifier-free formulas
produced by using Reddy and Loveland's quantifier elimination method.
From this analysis, we obtain the upper bound on the size of the
minimal deterministic WA for PA formulas.

We also show that the upper bound on the size of deterministic WAs for
formulas is tight. In fact, we show a stronger result. Namely, we give
a family of Presburger arithmetic formulas for which even a
nondeterministic WA must have at least triple exponentially many
states.

Furthermore, we investigate the automata constructed from atomic
formulas.  Specific algorithms for constructing WAs for linear
(in)equa\-tions have been developed
in~\cite{Boudet_Comon.1996,Boigelot.1999,Wolper_Boigelot.2000,Bartzis_Bultan.2003,Ganesh_Berezin_Dill.2002}\@.
We give upper and lower bounds on the automata size for linear
(in)equations and we improve the automata constructions
in~\cite{Boigelot.1999,Wolper_Boigelot.2000,Ganesh_Berezin_Dill.2002}
for linear (in)equations. We prove that our automata constructions are
optimal in the sense that the constructed deterministic WAs are
minimal.

We proceed as follows.  In~\S\ref{PRELIM}, we give background.
In~\S\ref{AUTO}, we investigate the WAs for quantifier-free formulas.
In~\S\ref{BOUND}, we prove the upper bound on the size of the minimal
deterministic WA for PA formulas and in~\S\ref{WORST}, we give a worst
case example.  Finally, in~\S\ref{CONCL}, we draw conclusions.

\section{Preliminaries}
\label{PRELIM}

\subsection{Presburger Arithmetic}

\emph{Presburger arithmetic} (PA) is the first-order logic over the
structure $\frakZ:=(\Int,<,+)$\@.  We use standard notation. For
instance, we write $\frakZ\models\varphi[a_1,\dots,a_r]$ for a formula
$\varphi(x_1,\dots,x_r)$ and $a_1,\dots,a_r\in\Int$ if $\varphi$ is
true in $\frakZ$ when the variable $x_i$ is interpreted as the integer
$a_i$, for $1\leq i\leq r$\@.  Analogously, $t[a_1,\dots,a_r]$ denotes
the integer when the $x_i$s are interpreted as the $a_i$s in the term
$t(x_1,\dots,x_r)$\@.  For a formula $\varphi(x_1,\dots,x_r)$, we
define $\Zrepresents{\varphi}:=
\set{(a_1,\dots,a_r)\in\Int^r}{\frakZ\models\varphi[a_1,\dots,a_r]}$\@.

\subsubsection{Extended Logical Language} 

We extend the logical language of PA by (i)~constants for the integers
$0$ and $1$, (ii)~the unary operation ``$-$'' for integer negation,
and (iii)~the unary predicates ``$d\divides$'' for the relation
``divisible by $d$,'' for each $d\geq 2$\@.  These constructs are
definable in PA, \eg, the formula $\exists x(x+\dots+x\EQ t)$ defines
$d\divides t$, where $x$ occurs $d$ times in the term $x+\dots+x$ and
$x$ does not appear in the term $t$\@.
The reason for the extended logical language, where~(i),~(ii),~and
(iii)~are treated as primitives, is that it admits quantifier
elimination, \ie, for a formula $\exists x\varphi(x,\overline{y})$,
where $\varphi$ is quantifier-free, we can construct a logically
equivalent quantifier-free formula $\psi(\overline{y})$\@.

Additionally, we allow the relation symbols $\leq,>,\geq$, and $\not=$
with their standard meanings.
In the following, we assume that terms and formulas are defined in
terms of the extended logical language for PA\@.  
We denote by $\pa$ the set of all Presburger arithmetic formulas over
the extended logical language and $\qf$ denotes the set of
quantifier-free formulas.

For convenience, we use standard symbols when writing terms.
For instance, $c$ stands for $1+\dots+1$ (repeated $c$ times) if
$c>0$, and $-(1+\dots+1)$ if $c<0$\@.  We call the term $c$ a
\emph{constant} and identify the term $c$ with the integer that it
represents.  Analogously, we write $k\CDOT x$ for $x+\dots+x$
(repeated $k$ times) if $k>0$, and $-(x+\dots+x)$ if $k<0$\@.
Moreover, if $k=0$ then $k\CDOT x$ abbreviates $x+(-x)$\@.  We say
that $k$ is a \emph{coefficient}.
For a term $t$ and $k\in\Int$, $k\cdot t$ denotes the term where the
constant and the coefficients in $t$ are multiplied by $k$\@.

A term $t$ is \emph{homogeneous} if it is either $0$ or of the form
$k_1\CDOT x_1+\dots+k_r\CDOT x_r$, for some $r\geq 1$, where the
variables $x_1,\dots,x_r$ are pairwise distinct and
$k_1,\dots,k_r\in\Int\setminus\{0\}$\@.
The \emph{normalized form} of $t_1\ZAECKZAECK t_2$, with
$\zaeckzaeck\in\{=,\not=,<,\leq,>,\geq\}$, is the logically equivalent
(in)equa\-tion $t\ZAECKZAECK c$, where summands of the form $k\CDOT x$
in $t_1$ and $t_2$ are collected on the left-hand side $t$ and
constants in $t_1$ and $t_2$ are collected on the right-hand side $c$
according to standard calculation rules. The \emph{normalized form} of
$d\divides t$ is the formula $d\divides t'+c$, where $c\in\Int$ is the
sum of the constants in $t$ and $t'$ is the homogeneous term in which
the coefficients of the summands of the form $k\CDOT x$ in $t$ are
collected.
We use $\Atoms(\varphi)$ to denote the set of atomic formulas
occurring in $\varphi\in\pa$ in their normalized forms.

\subsubsection{Formula Length}

The \emph{length of a formula} is the number of letters used in
writing the formula.  Note that the length of a formula depends
significantly on how we define the length of coefficients and
constants.  For instance, $x\EQ 10\CDOT y$ contains $6$ letters,
namely, $x$, $\EQ$, $1$, $0$, $\CDOT$, and $y$\@.  The ``expanded
version'' has $2+19$ letters since $10\CDOT y$ abbreviates the term
$y+y+y+y+y+y+y+y+y+y$\@.  We use the same definition of the length of
a formula as
in~\cite{Oppen.1978,Fischer_Rabin.1974,Reddy_Loveland.1978}\@. In
particular, the length of a coefficient or constant is the number of
letters of the expanded version.
However, it is possible to express $k\CDOT x$ by a formula of length
$\BigOh(\log\abs{k})$\@.  The idea is illustrated by $x\EQ 10\CDOT y$:
the formula is logically equivalent to $\exists z(x\EQ z+z\AND\exists
x(z\EQ x+x+y\AND x\EQ y+y))$\@. Note that we only need a fixed number
of variables for any $k$ (see~\cite{Fischer_Rabin.1974})\@.  For the
sake of uniformity, we define the length of the formula $d\divides t$
as the length of the term $t$ plus $d+1$\@.  Again, there is a
logically equivalent formula of length $\BigOh(\log d)$ plus the length
of $t$\@.  For the results in this paper it does not matter if we
define the length of an integer $k$ as $\BigOh(\log\abs{k})$ or as
$\BigOh(\abs{k})$\@.

\subsubsection{Nesting of Quantifiers}

\newcommand{\qn}{\operatorname{qn}}
\newcommand{\qa}{\operatorname{qa}}
\newcommand{\qbl}{\operatorname{qbl}}

It is well-known that we obtain coarse complexity bounds for checking
satisfiability if we only take into account the formula length.  We
obtain more precise complexity bounds when we additionally for account
the number of quantifiers and the number of quantifier alternations.

The \emph{quantifier number} of $\varphi\in\pa$ is the number of
quantifiers occurring in $\varphi$, \ie,
\begin{equation*}
  \qn(\varphi):=\begin{cases}
    \qn(\psi) & \text{if $\varphi=\neg\psi$,}
    \\
    \qn(\psi_1)+\qn(\psi_2) & \text{if $\varphi=\psi_1\oplus\psi_2$
      with $\oplus\in\{\wedge,\vee,\rightarrow,\leftrightarrow\}$,}
    \\
    1+\qn(\psi) & \text{if $\varphi=Qx\psi$ with
      $Q\in\{\exists,\forall\}$,}
    \\
    0 & \text{otherwise.}
  \end{cases}
\end{equation*}
For a quantifier $Q\in\{\exists,\forall\}$, $\overline{Q}$ denotes its
dual, \ie, $\overline{Q}:=\forall$ if $Q=\exists$, and
$\overline{Q}:=\exists$ if $Q=\forall$\@.
The number of \emph{quantifier alternations} of $\varphi\in\pa$ is
\begin{equation*}
  \qa(\varphi):=
  \min\{\qa_{\exists}(\varphi),\qa_{\forall}(\varphi)\}
  \,,
\end{equation*}
where
\begin{equation*}
  \qa_Q(\varphi):=\begin{cases}
    \qa_{\overline{Q}}(\psi) & \text{if $\varphi=\neg\psi$,}
    \\
    \max\{\qa_Q(\psi_1),\qa_Q(\psi_2)\} & \text{if
      $\varphi=\psi_1\oplus\psi_2$ with $\oplus\in\{\vee,\wedge\}$,}
    \\
    \qa_Q(\neg\psi_1\vee\psi_2)
    &\text{if $\varphi=\psi_1\rightarrow\psi_2$,}
    \\
    \qa_Q((\psi_1\rightarrow\psi_2)\wedge(\psi_2\rightarrow\psi_1))
    & \text{if $\varphi=\psi_1\leftrightarrow\psi_2$,}
    \\
    1+\qa_{\overline{Q}}(\psi) & \text{if $\varphi=\overline{Q}x\psi$,}
    \\
    \max\{1,\qa_Q(\psi)\} & \text{if $\varphi=Qx\psi$,}
    \\
    0 & \text{otherwise,}
  \end{cases}
\end{equation*}
for $Q\in\{\exists,\forall\}$\@.


\subsection{Automata over Finite Words}

The set of all words over an alphabet $\Sigma$ is denoted by
$\Sigma^*$, $\Sigma^+$ denotes the set of all non-empty words over
$\Sigma^*$, and $\lambda$ denotes the \emph{empty word}. The
\emph{length of the word} $w\in\Sigma^*$ is denoted by $\length{w}$\@.

A \emph{deterministic word automaton} (DWA) is a tuple
$\autA=(Q,\Sigma,\delta,q_{\rmI},F)$, where $Q$ is a finite set of
states, $\Sigma$ is a finite alphabet,
$\delta:Q\times\Sigma\rightarrow Q$ is the transition function,
$q_{\rmI}\in Q$ is the initial state, and $F\subseteq Q$ is the set of
accepting states. The \emph{size} of $\autA$ is the cardinality of
$Q$\@.  The \emph{language} of $\autA$ is
$L(\autA):=\set{w\in\Sigma^*}{\widehat{\delta}(q_{\rmI},w)\in F}$,
where $\widehat{\delta}(q,\lambda):=q$ and
$\widehat{\delta}(q,wb):=\delta(\widehat{\delta}(q,w),b)$, for $q\in
Q$, $b\in\Sigma$, and $w\in\Sigma^*$\@.
A state $q\in Q$ is \emph{reachable} from $p\in Q$ if there is a word
$w\in\Sigma^*$ such that $\widehat{\delta}(p,w)=q$\@.

Let $\autA=(Q,\Sigma,\delta,q_{\rmI},F)$ be a DWA, where we assume
that every state is reachable from $q_{\rmI}$\@. Note that the states
that are not reachable from $q_{\rmI}$ have no affect on the language
of the DWA and can be eliminated.
The states $p,q\in Q$ are \emph{equivalent}, $p\sim_{\autA} q$ for
short, if for all $w\in\Sigma^*$, we have that
$\widehat{\delta}(p,w)\in F$ iff $\widehat{\delta}(q,w)\in F$\@.  We
omit the subscript in the relation $\sim_{\autA}$ if $\autA$ is clear
from the context.
Note that $\mathbin{\sim}\subseteq Q\times Q$ is an equivalence
relation.  We denote the equivalence class of $q\in Q$ by
$\widetilde{q}$\@.
Since we assume that all states are reachable from $q_{\rmI}$, the
states $p,q\in Q$ can be merged iff $p\sim q$\@. We obtain the DWA
$\widetilde{\autA}:=(\set{\widetilde{q}}{q\in
  Q},\Sigma,\widetilde{\delta},
\widetilde{q_{\rmI}},\set{\widetilde{q}}{q\in F})$ with
$\widetilde{\delta}(\widetilde{q},b):= \widetilde{\delta(q,b)}$, for
$q\in Q$ and $b\in\Sigma$\@.  We have that
$L(\widetilde{\autA})=L(\autA)$ and $\widetilde{\autA}$ is
\emph{minimal}, \ie, for every DWA $\autB$ with $L(\autB)=L(\autA)$,
either $\autB$ has more states than $\widetilde{\autA}$ or $\autB$ is
isomorphic to $\widetilde{\autA}$\@.

\section{Automata Constructions}
\label{AUTO}

In this section, we investigate the automata for quantifier-free PA
formulas.  In~\S\ref{subsec:encoding}, we define how DWAs recognize
sets of integers, in~\S\ref{subsec:(in)equations}, we provide optimal
automata constructions for linear (in)equa\-tions,
in~\S\ref{subsec:divisibility}, we give an automata construction for
the divisibility relation, and finally, in~\S\ref{subsec:qf_formulas},
we give an upper bound on the size of the minimal DWA for a
quantifier-free formula.

\subsection{Representing Sets of Integers with Automata}
\label{subsec:encoding}

We use an idea that goes back at least to
B{\"u}chi~\citeyear{Buechi.1960} for using automata to recognize
tuples of numbers by mapping words to tuples of numbers.  There are
many possibilities to represent integers as words.  We use an encoding
similar to~\cite{Boigelot.1999,Wolper_Boigelot.2000}, which is based
on the $\varrho$'s complement representation of integers, where
$\varrho\geq 2$ and the most significant bit is the first digit.
For the remainder of the
paper, we fix $\varrho\geq 2$ and let $\Sigma$ be the alphabet
$\{0,\dots,\varrho-1\}$\@.
\begin{definition}
  For $b_{n-1}\dots b_0\in\Sigma^*$, we define $\tonats{b_{n-1}\dots
    b_0}:=\sum_{0\leq i<n} \varrho^i b_i$\@.  We generalize this
  encoding to integers as follows.  For $b_nb_{n-1}\dots
  b_0\in\Sigma^+$, we define
  \begin{equation*}
    \toints{b_nb_{n-1}\dots b_0}:=
    \tonats{b_{n-1}\dots b_0}
    -
    \begin{cases}
      0 & \text{if $b_n=0$,}
      \\
      \varrho^{n} & \text{if $b_n\not=0$\@.}
    \end{cases}
  \end{equation*}
  We call the first letter $b_n$ the \emph{sign letter}, since it
  determines whether the word represents a positive or a negative
  number.
\end{definition}
Note that the empty word $\lambda$ does not represent an integer.
This requirement saves us from considering some special cases in
\S\ref{subsubsec:opt_inequations} and
\S\ref{subsubsec:opt_inequations} where we optimize the automata
constructions for (in)equations.  However, for the natural numbers, it
holds that $\tonats{\lambda}=0$\@.  Furthermore, note that the
encoding of an integer is not unique.
First, we have that $\toints{bu}=\toints{bcu}$, where $b,c\in\Sigma$
and $u\in\Sigma^*$ with $c=0$ if $b=0$ and $c=\varrho-1$, otherwise.
Second, it holds that $\toints{bu}=\toints{b'u}$, for all
$u\in\Sigma^*$ and $b,b'\in\Sigma\setminus\{0\}$, \ie, the sign letter
$b\not=0$ can be replaced by any other letter $b'\not=0$\@.  The
motivation for allowing any letter to be the sign letter is that we do
not have to deal with words in $\Sigma^+$ that do not represent an
integer. This eliminates case distinctions of the automata
constructions in the next subsections.

We extend the encoding to tuples of natural numbers and integers as
follows: A word
$w:=\overline{b}_{n-1}\dots\overline{b}_0\in(\Sigma^r)^*$
\emph{represents} the tuple $\overline{a}:=(a_1,\dots,a_r)\in\Nat^r$
of integers, where the $i$th ``track'' of the word $w$ encodes the
natural number $a_i$\@. That is, for all $1\leq i\leq r$, we have that
$a_i=\toints{b_{n-1,i}\dots b_{0,i}}$, where
$\overline{b}_j=(b_{j,1},\dots,b_{j,r})$ for $0\leq j<n$\@.  
The encoding of an integer tuple
$\overline{z}=(z_1,\dots,z_r)\in\Int^r$ is defined analogously for a
word
$w=\overline{b}_n\overline{b}_{n-1}\dots\overline{b}_0\in(\Sigma^r)^+$\@.
The first letter $\overline{b}_n$ of $w$ is the \emph{sign letter}
since it determines the signs of the integers $z_1,\dots,z_r$\@.  We
define $\sigma(\overline{b}_n):=(c_1,\dots,c_r)$, where $c_i=0$ if the
$i$th coordinate of $\overline{b}_n$ is $0$ and $c_i=-1$, otherwise,
for each $1\leq i\leq r$\@.
We abuse notation and write $\tonats{w}$ to denote the tuple
$\overline{a}\in\Nat^r$ and $\toints{w}$ to denote the integer tuple
$\overline{z}$\@.

Moreover, we write $\towordsnats{\overline{a}}$ for the shortest word
in $(\Sigma^r)^*$ that represents $\overline{a}\in\Nat^r$\@. Note that
$\towordsnats{\overline{a}}$ is well-defined since (1)~there is a word
$w\in(\Sigma^r)^*$ with $\toints{w}=\overline{a}$, and (2)~if
$\tonats{v}=\tonats{v'}$ for $v,v'\in(\Sigma^r)^*$, then $v$ and $v'$
have a common suffix $u\in(\Sigma^r)^*$ with
$\tonats{u}=\tonats{v}$\@.
Similar to $\towordsnats{\overline{a}}$ for $\overline{a}\in\Nat^r$,
we define $\towordsints{\overline{z}}$, for $\overline{z}\in\Int^r$,
as the shortest word $w\in(\Sigma^r)^+$ with $\overline{z}=\toints{w}$
and the first letter of $w$ is in $\{0,\varrho-1\}^r$\@.

\begin{definition}
  Let $U\subseteq\Int^r$\@.  The language $L\subseteq (\Sigma^r)^*$
  \emph{represents} $U$ if $L=\set{w\in(\Sigma^r)^+}{\toints{w}\in
    U}$\@.
  A DWA $\autA$ \emph{represents} $U$ if $L(\autA)$ represents $U$\@.
\end{definition}
Note that by this definition not every language over $\Sigma^r$
represents a set of tuples of integers, and not every DWA with
alphabet $\Sigma^r$ represents a subset of $\Int^r$\@.
%
\begin{example}
  The set of pairs $(x,y)\in\Int^2$ where $y$ equals $2x$ is
  represented by the DWA depicted in
  Figure~\ref{fig:ex_representing_integers} by using the base
  $\varrho=2$ for representing integers as words, \ie, the alphabet of
  the DWA is $\{0,1\}^2$\@.  In the figure, we use abbreviations like
  $(0,\textrm{--})$ to denote the letters $(0,0)$ and $(0,1)$\@.
  \begin{figure}[t]
    \centering
    \includegraphics{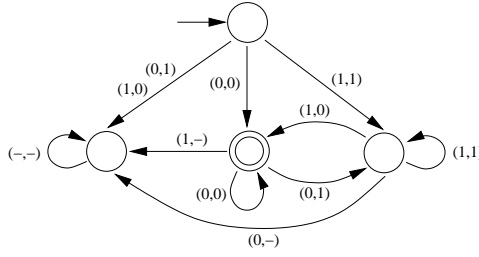}
    \caption{DWA over the alphabet $\{0,1\}^2$ representing the set
      $\set{(x,y)\in\Int^2}{y=2x}$\@.
      \label{fig:ex_representing_integers}}
  \end{figure}
\end{example}

\subsection{Linear Equations and Inequations}
\label{subsec:(in)equations}

In this subsection, we first recall the automata constructions given
in~\cite{Boigelot_Rassart_Wolper.1998,Boigelot.1999,Wolper_Boigelot.2000,Ganesh_Berezin_Dill.2002}
for linear (in)equa\-tions.  Then, we improve these constructions such
that they are optimal, \ie, the constructed DWAs are minimal.  Assume
that the (in)equa\-tion $t\ZAECKZAECK c$ is given in normalized form,
\ie, $t(x_1,\dots,x_r)$ is a homogeneous term,
$\zaeckzaeck\in\{=,\not=,<,\leq,>,\geq\}$, and $c\in\Int$\@.

First, we make the following observation for a word $u\in(\Sigma^r)^*$
and $\overline{b}\in\Sigma^r$\@.  If $u\not=\lambda$ then
$\toints{u\overline{b}}=\varrho\toints{u}+\overline{b}$\@. For
$u=\lambda$, we have that
$\toints{\overline{b}}=\sigma(\overline{b})$\@.
Given this, it is relatively straightforward to obtain an analog of a
DWA with \emph{infinitely} many states for $t\ZAECKZAECK c$\@. The set
of states is $\{q_{\rmI}\}\cup\Int$, where $q_{\rmI}$ is the initial
state.  Note that we identify integers with states.  The idea is to
keep track of the value of $t$ as successive bits are read.  Thus,
except for the special initial state, a state in $\Int$ represents the
current value of $t$\@.
Lemma~\ref{lem:automaton_infinite_(in)equation} below justifies this
intuition.  The transition function
$\eta:(\{q_{\rmI}\}\cup\Int)\times\Sigma^r\rightarrow(\{q_{\rmI}\}\cup\Int)$
is defined as follows for a letter $\overline{b}\in\Sigma^r$\@.  For
the initial state, we define
$\eta(q_{\rmI},\overline{b}):=t[\sigma(\overline{b})]$\@.  For
$q\in\Int$, we define $\eta(q,\overline{b}):= \varrho
q+t[\overline{b}]$\@.
\begin{lemma}
  \label{lem:automaton_infinite_(in)equation}
  For $u\in(\Sigma^r)^*$ of length $n\geq 0$ we have that
  \begin{enumerate}[(a)]
  \item $\widehat{\eta}(q,u)= \varrho^{n}q+t\bigl[\tonats{u}\bigr]$,
    for $q\in\Int$, and
  \item $\widehat{\eta}(q_{\rmI},\overline{b}u)=
    t\bigl[\toints{\overline{b}u}\bigr]$, for
    $\overline{b}\in\Sigma^r$\@.
  \end{enumerate}
\end{lemma}
\begin{proof}
  (a)~is easily proved by induction over $n$, and (b)~follows from~(a)
  and the definition of $\eta$\@.
\end{proof}

Later we make use of the following lemma, which translates the
question whether $q\in\Int$ is reachable from $p\in\Int$ via
$\widehat{\eta}$ to a number-theoretic problem.
\begin{lemma}
  \label{lem:reachability_question}
  Let $p,q\in\Int$\@. There are $N,a_1,\dots,a_r\geq 0$ such that
  $N\geq\lceil\log_{\varrho}(1+\max\{a_1,\dots,a_r\})\rceil$ and
  $\varrho^N p+t[a_1,\dots,a_r]=q$ iff there is a word
  $w\in(\Sigma^r)^*$ such that $\widehat{\eta}(p,w)=q$\@.
\end{lemma}
\begin{proof}
  ($\Rightarrow$)\ 
  Assume that $\towordsnats{a_1,\dots,a_r}$ has length $\ell$\@.  Note
  that $\ell\leq N$\@. This follows from the fact that for every
  $a\in\Nat$, there is a word $u\in\Sigma^*$ of length
  $\lceil\log_{\varrho}(1+a)\rceil$ such that $\tonats{u}=a$\@.  By
  Lemma~\ref{lem:automaton_infinite_(in)equation}(a), we have that
  \begin{equation*}
    \widehat{\eta}
    \bigl(p,\overline{0}^{N-\ell}\towordsnats{a_1,\dots,a_r}\bigr)
    =\varrho^N p+t[a_1,\dots,a_r]=q
    \,.
  \end{equation*}

  \vspace{\topsep}
  \noindent
  ($\Leftarrow)$\ 
  Assume that $\widehat{\eta}(p,w)=q$, for some $w\in(\Sigma^r)^*$\@.
  Let $N$ be the length of $w$\@.  We have that $N\geq
  {\lceil\log_{\varrho}(1+a)\rceil}$, where $a$ is the largest number
  in the tuple $\tonats{w}$\@. 
  It follows from Lemma~\ref{lem:automaton_infinite_(in)equation}(a)
  that $\widehat{\eta}(p,w)=\varrho^N p+t[\tonats{w}]$\@.
\end{proof}

The automata constructions
in~\cite{Wolper_Boigelot.2000,Ganesh_Berezin_Dill.2002} are based on
the observation that the states $q,q'\in\Int$ can be merged if,
intuitively speaking, $q$ and $q'$ are both small or both large.
Here, the meaning of ``small'' and ``large'' depends on the
coefficients of $t$ and on the constant $c$\@.  More precisely, we say
that $q\in\Int$ is \emph{small} if $q<\min\{c,-\normpos{t}\}$, and
\emph{large} if $q>\max\{c,\normneg{t}\}$, where 
\begin{equation*}
  \normneg{t}:=
  \sum_{\substack{1\leq j\leq r\\\text{and }k_j<0}} \abs{k_j}
  \qquad\text{and}\qquad
  \normpos{t}:=
  \sum_{\substack{1\leq j\leq r\\\text{and }k_j>0}} k_j
\end{equation*}
assuming that $t$ is of the form $k_1\CDOT x_1+\dots+k_r\CDOT x_r$\@.
%
Note that from a small value we can only obtain smaller values and
from a large value we can only obtain larger values by $\eta$, \ie,
for all $\overline{b}\in\Sigma^r$, if $q>\normneg{t}$ then
$\eta(q,\overline{b})=\varrho q+t[\overline{b}]>q$, and if
$q<-\normpos{t}$ then $\eta(q,\overline{b})=\varrho
q+t[\overline{b}]<q$\@.
A difference between the constructions in~\cite{Wolper_Boigelot.2000}
and~\cite{Ganesh_Berezin_Dill.2002} are the bounds that determine the
meaning of ``small'' and ``large''.

For $m<n$, we define $\autA^{t\ZAECKZAECK
  c}_{(m,n)}:=(Q,\Sigma^r,\delta,q_{\rmI},F)$, where
$Q:=\{q_{\rmI}\}\cup\set{q\in\Int}{m\leq q\leq n}$ and
\begin{equation*}
  \delta(q,\overline{b}):=\begin{cases}
    m & \text{if $\eta(q,\overline{b})\leq m$,}\\
    n & \text{if $\eta(q,\overline{b})\geq n$,}\\
    \eta(q,\overline{b}) & \text{otherwise,}
  \end{cases}
\end{equation*}
for $q\in Q$ and $\overline{b}\in\Sigma^r$\@.  Moreover, let
$F:=\set{q\in Q\cap\Int}{q\zaeckzaeck c}$\@.
\begin{lemma}
  \label{fact:bound_automata_(in)equation}
  The DWA $\autA^{t\ZAECKZAECK c}_{(m,n)}$ represents
  $\Zrepresents{t\ZAECKZAECK c}$ if $m$ is small and $n$ is large.
  Moreover, $\autA^{t\ZAECKZAECK c}_{(m,n)}$ has $2+n-m$ states.
\end{lemma}
\begin{proof}
  The fact that $\autA^{t\ZAECKZAECK c}_{(m,n)}$ represents
  $\Zrepresents{t\ZAECKZAECK c}$ follows from
  Lemma~\ref{lem:automaton_infinite_(in)equation}, and
  $\autA^{t\ZAECKZAECK c}_{(m,n)}$ has $2+n-m$ states by definition.
\end{proof}

In the following, we optimize the constructions such that the produced
DWA for an (in)equa\-tion is minimal.  Moreover, we give a lower bound
on the minimal DWA for an (in)equa\-tion. However, these results are
not needed for the upper bound on the minimal DWA for a PA formula.
In the remainder of this subsection, let $\autA^{t\ZAECKZAECK
  c}_{(m,n)}= (Q,\Sigma^r,\delta,q_{\rmI},F)$ for the (in)equa\-tion
$t\ZAECKZAECK c$ with $m=\max\set{q\in\Int}{q\text{ is small}} $ and
$n=\min\set{q\in\Int}{q\text{ is large}}$\@.  We restrict ourselves to
the cases where $\zaeckzaeck\in\{=,<,>\}$\@.
The cases with $\zaeckzaeck\in\{\not=,\leq,\geq\}$ reduce to the cases
for $=$, $<$, $>$ and complementation of DWAs, since $t\NOTEQ c$ is
logically equivalent to $\NEG t\EQ c$, $t\LEQ c$ is logically
equivalent to $\NEG t\GREATER c$, and $t\GEQ c$ is logically
equivalent to $\NEG t\LESS c$\@.  Note that complementation of a DWA
can be done by flipping accepting and non-accepting states.  After
complementation we have to make the initial state of the DWA
non-accepting since the empty word does not represent any integer
tuple.  The resulting DWA is minimal iff the original DWA is minimal.

\subsubsection{Eliminating Unreachable States}

An obvious optimization is to eliminate the states in $Q\cap\Int$ that
are not a multiple of the greatest common divisor of the absolute
values of the coefficients in the term $t$, since they are not
reachable from the initial state $q_{\rmI}$\@. We define the
\emph{greatest common divisor} of the term $t(x_1,\dots,x_r)$ as
$\gcd(t):=\gcd(\abs{k_1},\dots,\abs{k_r})$, where $k_i$ is the
coefficient of the variable $x_i$, for $1\leq i\leq r$\@.
\begin{lemma}
  \label{lem:reachable}
  The state $q\in Q\cap\Int$ is reachable from the initial state
  $q_{\rmI}$ iff $q$ is a multiple of $\gcd(t)$\@.
\end{lemma}
\begin{proof}
  ($\Rightarrow$)\ %
  This direction is easy to prove by induction on the length of
  $w\in(\Sigma^r)^*$ with $\widehat{\delta}(q_{\rmI},w)\in\Int$: for
  all $\overline{b}\in\Sigma^r$, it holds that (i)
  $\delta(q_{\rmI},\overline{b})=t[\sigma(\overline{b})]$ is a
  multiple of $\gcd(t)$, and (ii) if
  $\widehat{\delta}(q_{\rmI},w)\in\Int$ is a multiple of $\gcd(t)$
  then $\varrho\widehat{\delta}(q_{\rmI},w) + t[\overline{b}]$ is a
  multiple of $\gcd(t)$\@.
  
  \vspace{\topsep}
  \noindent
  ($\Leftarrow$)\ %
  Assume that $q$ is a multiple of $\gcd(t)$\@.  There are
  $v_1,\dots,v_r\in\Int$ such that $t[v_1,\dots,v_r]=q$\@.  With
  Lemma~\ref{lem:automaton_infinite_(in)equation}(b) we conclude that
  $\widehat{\delta}\bigl(q_{\rmI},\towordsints{v_1,\dots,v_r}\bigr)=
  t[v_1,\dots,v_r]$\@.
\end{proof}

Alternatively, instead of filtering out the states $q\in\Int$ that are
not a multiple of $\gcd(t)$ we can rewrite the (in)equa\-tion
$t\ZAECKZAECK c$ to the logically equivalent atomic formula $\alpha$
and then construct the DWA for $\alpha$, where $\alpha$ is defined as
\begin{equation*}
  \alpha:=\begin{cases}
    t'\ZAECKZAECK \bigl\lceil \frac{c}{\gcd(t)}\bigr\rceil 
    & \text{if $\zaeckzaeck$ is $<$,}
    \\
    t'\ZAECKZAECK\bigl\lfloor \frac{c}{\gcd(t)}\bigr\rfloor & 
    \text{if $\zaeckzaeck$ is $>$,}
    \\
    t'\ZAECKZAECK\frac{c}{\gcd(t)} & \text{if $\zaeckzaeck$ is $=$
      and  $c$ is a multiple of $\gcd(t)$,}
    \\
    1<0 & \text{otherwise,}
  \end{cases}
\end{equation*}
where the coefficients in $t'$ are the coefficients of $t$ divided
by $\gcd(t)$\@.  
In the remainder of this subsection we assume that $\gcd(t)=1$\@.

\subsubsection{Optimal Construction for Inequations} 
\label{subsubsec:opt_inequations}

In the following, we assume that the inequation is of the form
$t\GREATER c$, with $c\geq 0$\@.  The cases where $\zaeckzaeck$ is $<$
or $c\geq 0$ are analogous.  The following example illustrates that
many states of $\autA^{t\GREATER c}_{(m,n)}$ can be merged if $c$ is
significantly larger than $\normneg{t}$\@.
\begin{example}
  \label{ex:minimal_inequation}
  The automata construction described above for the inequation
  $x-y\GREATER 32$ produces a DWA with the set of states
  $Q=\{q_{\rmI},-2,-1,0,\dots,32,33\}$; but the minimal DWA (see
  Figure~\ref{fig:minimal_automaton_inequation}) for $x-y>32$ has only
  $13$ states when we choose the base $\varrho=2$\@.
  \begin{figure}[t]
    \centering
    \includegraphics{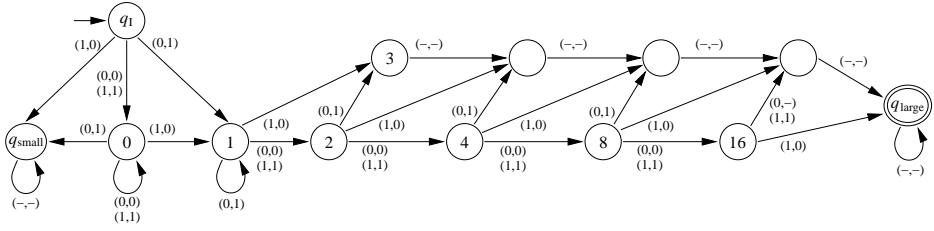}
    \caption{Minimal DWA over the alphabet $\{0,1\}^2$ for the inequation 
      $x-y>32$\@.
      \label{fig:minimal_automaton_inequation}}
  \end{figure}
  
  The reason for this gap is that several states can be merged.
  First, we merge the states $-2$ and $-1$ since from both states only
  non-accepting states are reachable.
  Second, we can merge the states in $Q':=\set{q\in
    Q\cap\Int}{2q+a-b>c, \text{ for all $a,b\in\{0,1\}$}}=
  \{17,\dots,32\}$ to a single state since all states in $Q'$ are
  non-accepting and all their transitions go to state $33$\@.  The
  state $16$ cannot be merged with any other state since if we read
  the letter $(1,0)$, we end up in the accepting state $33$, and if we
  read the letters $(0,0)$, $(1,1)$, or $(0,1)$ we end up in the
  non-accepting states $32$ or $31$\@.  The states in $\{9,\dots,15\}$
  can again be merged to a single state since with every transition we
  reach a state in $Q'$\@. Analogously, we can merge the states in
  $\{5,6,7\}$\@.
\end{example}  

In the following, we determine the equivalent states in
$\autA^{t\GREATER c}_{(m,n)}$. Note that from
Lemma~\ref{lem:reachable} it follows that all states are reachable
from $q_{\rmI}$ since we assume that $\gcd(t)=1$\@.  We use the
notation $[d,d')$ for the set $\{d,\dots,d'-1\}$ if $d,d'\in\Int$, and
if $d\in\Int$ and $d'=\infty$ then $[d,d'):=\set{z\in\Int}{z\geq
  d}$\@.
In order to identify the equivalent states, we define the following
strictly monotonically decreasing sequence $d_0>d_1>\dots>d_{\ell}$,
for some $\ell\geq 1$\@.  Let $d_0:=\infty$ and
$d_1:=\max\{c+1,\normneg{t}\}$\@.  Assume that $d_0>d_1>\dots>d_i$ are
already defined, for some $i\geq 1$\@.
\begin{itemize}
\item If $d_i=\normneg{t}$ then we are done, \ie, $\ell=i$\@.
\item If $d_i>\normneg{t}$ then let $d_{i+1}<d_i$ be the smallest
  integer greater than $\normneg{t}-1$ such that for all
  $\overline{b}\in\Sigma^r$, there is an index $j$ with $1\leq j\leq
  i$ and
  \begin{equation}
    \label{eqn:interval}
    \varrho d_{i+1}+t[\overline{b}],\,\,
    \varrho (d_i-1)+t[\overline{b}]
    \,\in\,[d_j,d_{j-1})
    \,.
  \end{equation}
  Note that $d_{i+1}$ is well-defined since $d_i-1$
  satisfies~(\ref{eqn:interval}), for all
  $\overline{b}\in\Sigma^r$\@.
\end{itemize}

The following lemma characterizes the equivalent states. In
particular, it shows that we can merge the states in
$R:=\{-\normpos{t}, \normpos{t}-1\}$, and for each $i\leq i\leq\ell$,
the states in $[d_i,d_{i-1})$ can be collapsed to one state.
\begin{lemma}
  \label{lem:equivalent_states}
  For all $p,q\in Q$, it holds that $p\sim q$ iff $p=q$ or $p,q\in R$
  or $p,q\in[d_i,d_{i-1})$, for $1\leq i\leq\ell$\@.
\end{lemma}
\begin{proof}
  ($\Leftarrow$)\ %
  If $p=q$ then it is obvious that $p\sim q$\@. If $p,q\in R$ then we
  also have that $p\sim q$, since both states are non-accepting and
  all transitions from these states either go to $-\normpos{t}$ or to
  $-\normpos{t}-1$\@.
  It remains to prove that for $1\leq i\leq\ell$, if
  $p,q\in[d_i,d_{i-1})$ then $p\sim q$\@.  We prove this claim by
  induction over $i$\@. For $i=1$, there is nothing to prove, since
  $[d_1,d_0)\cap Q$ is a singleton.  For the induction step, assume
  that $i>1$ and let $p,q\in[d_i,d_{i-1})$\@.  Without loss of
  generality we assume that $p\leq q$\@.  By the definition of the
  transition function $\delta$ and the sequence
  $d_0>d_1>\dots>d_{\ell}$, we have that
  \begin{equation*}
    \varrho d_i+t[\overline{b}]\leq 
    \delta(p,\overline{b})\leq
    \delta(q,\overline{b})\leq
    \varrho (d_{i-1}-1)+t[\overline{b}]
    \,,
  \end{equation*}
  for all $\overline{b}\in\Sigma^r$\@.  Since there is a $1\leq j<i$
  with $\varrho d_i+t[\overline{b}], \varrho
  d_{i-1}+t[\overline{b}]\in[d_j,d_{j-1})$ we conclude that
  $\delta(p,\overline{b}),\delta(q,\overline{b})\in[d_j,d_{j-1})$\@.
  The claim now follows from the induction hypothesis.

  \vspace{\topsep}
  \noindent
  ($\Rightarrow$)\ %
  We prove the claim by contraposition, \ie, $p\not\sim q$ is implied
  by the three conditions (i)~$p\not=q$, (ii)~$p\in R\Rightarrow
  q\not\in R$, and (iii)~for all $1\leq i\leq \ell$, $p\in
  [d_i,d_{i-1})\Rightarrow q\not\in [d_i,d_{i-1})$\@.
  Assume $p\not=q$\@.  It suffices to distinguish the following three
  cases.

  \newdef{case1}{Case}
  \begin{case1}[{\rm\!\!: $p\in R$ and $q\not\in R$}]
    Since we can reach an accepting state from $q$, we have that
    $p\not\sim q$\@.
  \end{case1}

  \begin{case1}[{\rm\!\!: $p\in[d_i,d_{i-1})$ and
      $q\not\in[d_i,d_{i-1})$, for some $1\leq i\leq\ell$}]  
    It is straightforward to prove by induction
    over $i$ that $p\not\sim q$\@.
  \end{case1}
  
  \begin{case1}[{\rm\!\!: $p\not\in R\cup\bigcup_{1\leq
        i\leq\ell}[d_i,d_{i-1})$}]
    Note that the conditions~(ii) and~(iii) are satisfied.  We have
    that either $p=q_{\rmI}$ or $p\in S$, where $S:=\set{s\in
      Q\cap\Int}{-\normpos{t}<s<\normneg{t}}$\@.
  
    If $p=q_{\rmI}$ and $q\in R$ then we conclude similar to Case~1
    that $p\not\sim q$\@.
    Assume that $p=q_{\rmI}$ and $q\not\in R$\@.  Let
    $\overline{b}\in\Sigma^r$ be the letter that has a $0$ in its
    $i$th coordinate iff the $i$th coefficient of $t$ is negative, and
    otherwise the $i$th coordinate is $\varrho-1$\@.  It holds that
    $q_{\rmI}\not\sim q$, since
    $\delta(q_{\rmI},\overline{b})=-t[\overline{b}]\in R$ and
    $\delta(q,\overline{b})=\varrho q+\varrho\normpos{t}\geq q$\@.
    From Case~1, it follows that $p\not\sim q$\@.
  
    Assume that $p\in S$\@.  Note that for every $s\in S$ there is a
    $\overline{b}\in\Sigma^r$ such that $\delta(s,\overline{b})\in
    S$\@.  It follows that for every $n\geq 0$ there is a word
    $u\in(\Sigma^r)^*$ of length $n$ such that
    $\widehat{\delta}(p,u)\in S$\@.  We conclude that there is a word
    $u\in(\Sigma^r)^*$ such that $\widehat{\delta}(p,u)\in S$ and
    $\widehat{\delta}(q,u)\in R \cup\bigcup_{1\leq
      i\leq\ell}[d_i,d_{i-1})$, since
    $\delta(s,\overline{b})-\delta(s',\overline{b})=\varrho(s-s')$,
    for all $s,s'\in S$ and all $\overline{b}\in\Sigma^r$\@.
    Analogously to the Cases~1 and~2 we conclude that $p\not\sim q$\@.
    \qed
  \end{case1}
\end{proof}

From Lemma~\ref{lem:equivalent_states}, it follows that the minimal
DWA representing $\Zrepresents{t\GREATER c}$ has at least
$\normneg{t}+\normpos{t}$ states.  Note that this is in contrast to
the number of symbols we need to write the inequation $t\GREATER c$ if
coefficients are represented as binary numbers. For instance, we need
$22+7$ letters for $1025\cdot x-1024\cdot y\GREATER 0$, since each of
the two coefficients can be represented with $11$ digits. The same
lower bound on the minimal DWA size holds for $t\LESS c$\@.  In the
following, we show that a similar lower bound holds for equations.

\subsubsection{Optimal Construction for Equations}
\label{subsubsec:opt_equations}

For an equation $t\EQ c$, we can collapse the states in $\autA^{t\EQ
  c}_{(m,n)}$ from which we cannot reach the accepting state $c\in Q$
to a single non-accepting state.
These optimizations produce the minimal DWA for $t\EQ c$\@.  For
instance, the case for $p\in Q\cap\Int$ is proved as follows. Assume
that we can reach the state $c$ from $p\in Q\cap\Int$, \ie, there is a
$u\in(\Sigma^r)^*$, with $\widehat{\delta}(p,u)=c$\@.  Any other
states $q\in Q\cap\Int$ with $q\not=p$ from which we can reach $c$
cannot be merged with $p$, since
\begin{equation*}
  c
  =
  \widehat{\delta}(p,u)
  \,\overset{\textup{Lemma~\ref{lem:automaton_infinite_(in)equation}(a)}}{=}\,
  \varrho^{\length{u}} p+t\bigl[\tonats{u}\bigr]\not=
  \varrho^{\length{u}} q+t\bigl[\tonats{u}\bigr]
  \,\overset{\textup{Lemma~\ref{lem:automaton_infinite_(in)equation}(a)}}{=}\,
  \widehat{\delta}(q,u)
  \,.
\end{equation*}
The other cases are proved similarly.

A lower bound for the minimal DWA representing $\Zrepresents{t\EQ c}$
is based on the following lemma about the states of the DWA
$\autA^{t\ZAECKZAECK c}_{(m,n)}=(Q,\Sigma^r,\delta,q_{\rmI},F)$, where
$\zaeckzaeck\in\{=,\not=,<,\leq,>,\geq\}$\@.  Let $S:=\set{s\in
  Q\cap\Int}{-\normpos{t}<s<\normneg{t}}$ and $[n]:=\{0,\dots,n-1\}$,
for $n\geq 0$\@.
\begin{lemma}
  \label{lem:S}
  Every $q\in Q\cap\Int$ is reachable from every $p\in S$\@.
\end{lemma}
\begin{proof}
  We need a result from number theory. Let $\gamma>0$ and let
  $c_1,\dots,c_{\gamma}$ be integers with $0<c_1<\dots<c_{\gamma}$ and
  $\gcd(c_1,\dots,c_{\gamma})=1$\@.  The \emph{Frobenius number}
  $G(c_1,\dots,c_{\gamma})$ is the greatest integer $z$ for which the
  linear equation $c_1\cdot x_1+\dots+c_{\gamma}\cdot x_{\gamma}\EQ z$
  has \emph{no} solution in the natural numbers. For $\gamma=1$, it
  trivially holds that $G(c_1)=-1$\@. For $\gamma>1$, the upper bound
  $G(c_1,\dots,c_{\gamma})\leq\frac{c_{\gamma}^2}{\gamma-1}$ was
  proved by~\citeN{Dixmier.1990}\@. It is straightforward to show that
  for all $\gamma>0$,
  \begin{equation}
    \label{eqn:Frobenius}
    G(c_1,\dots,c_{\gamma})
    <
    \varrho^{c_1+\dots+c_{\gamma}}-(c_1+\dots+c_{\gamma})
    \,.
  \end{equation}
  
  In the following, we will prove the lemma, \ie, for $p\in S$ and
  $q\in Q\cap\Int$ there is a word $u\in(\Sigma^r)^*$ such that
  $\widehat{\delta}(p,u)=q$\@.  Note that if $r=0$ and $r=1$ then
  $S=\emptyset$ and the claim is trivially true.  Assume that $r\geq
  2$\@.  By Lemma~\ref{lem:reachability_question}, it suffices to show
  that the equation
  \begin{equation}
    \label{eqn:claim}
    \varrho^Np+t(x_1,\dots,x_r)=q
  \end{equation}
  has a solution $a_1,\dots,a_r\geq 0$ with
  $N\geq\lceil\log_{\varrho}(1+\max\{a_1,\dots,a_r\})\rceil$\@.  We
  distinguish four cases depending on $p$ and $q$\@.
  
  \newdef{case2}{Case}
  \begin{case2}[{\rm\!\!: $p=0$}]
    Equation~(\ref{eqn:claim}) simplifies to
    \begin{equation}
      \label{eqn:case_0}
      t(x_1,\dots,x_r)=q
      \,.
    \end{equation}
    There are positive and negative coefficients in $t$, since $p\in
    S$\@. It follows that equation~(\ref{eqn:case_0}) has infinitely
    many solutions in the natural numbers.  Recall that we assume that
    $\gcd(t)=1$\@.  In particular, there are $a_1,\dots,a_r\geq 0$
    with $\varrho^Np+t[a_1,\dots,a_r]=q$, for some appropriate large
    enough $N$\@.
  \end{case2}

  \begin{case2}[{\rm\!\!: $p>0$ and $q\geq 0$}]
    Let $k_{i_1},\dots,k_{i_{\mu}}$ be the positive coefficients in
    $t$, and let $k_{j_1},\dots,k_{j_{\nu}}$ be the negative
    coefficients in $t$\@.
    Let $N$ be the size of the DWA $\autA^{t\ZAECKZAECK c}_{(m,n)}$,
    \ie, $N=3+\max\{|c|,\normpos{t}\}+\max\{c,\normneg{t}\}$\@.  We
    rewrite equation~(\ref{eqn:claim}) to
    \begin{equation}
      \label{eqn:case_++}
      \varrho^Np-q+
      t_1(x_{i_1},\dots, x_{i_{\mu}})
      =
      t_2(x_{j_1},\dots, x_{j_{\nu}})
      \,,
    \end{equation}
    where $t_1$ is the term $k_{i_1}\CDOT
    x_{i_1}+\dots+k_{i_{\mu}}\CDOT x_{i_{\mu}}$, and $t_2$ is the term
    $|k_{j_1}|\cdot x_{j_1}+\dots+|k_{j_{\nu}}|\cdot x_{j_{\nu}}$\@.
    Note that $\varrho^Np-q\geq0$ since $p>0$ and $\varrho^N\geq q$\@.
    Let $D:=\gcd(|k_{j_1}|,\dots,|k_{j_{\nu}}|)$\@.  In order to show
    the existence of a solution $a_1,\dots,a_r\in[\varrho^N]$ of
    equation~(\ref{eqn:case_++}), we proceed in two steps:
    \begin{itemize}
    \item[\bf Step 1:] There are $a_{i_1},\dots,a_{i_{\mu}}\in[D]$ such
      that
      \begin{equation*}
        D
        \DIVIDES
        \varrho^Np-q+t_1[a_{i_1},\dots,a_{i_{\mu}}]
        \,.
      \end{equation*}
    \item[\bf Step 2:] There are $a_{j_1},\dots,a_{j_{\nu}}\in[\varrho^N]$ such
      that
      \begin{equation*}
        \varrho^Np-q+t_1[a_{i_1},\dots,a_{i_{\mu}}]
        =
        t_2[a_{j_1},\dots,a_{j_{\nu}}]
        \,.
      \end{equation*}
    \end{itemize}

    \vspace{\topsep}
    \noindent
    \textbf{Proof of Step~1:}
    If $\mu=0$ then there is nothing to prove. Assume that $\mu>0$\@.
    There are $K,R\geq 0$ such that $\varrho^Np-q=DK+R$ with $R<D$\@.
    It suffices to show that there are $a_{i_1},\dots,a_{i_{\mu}}$
    with $0\leq a_{i_1},\dots,a_{i_{\mu}}<D$, and $K'\geq 0$, such
    that $DK'=R+t_1[a_{i_1},\dots,a_{i_{\mu}}]$, since then
    \begin{equation*}
      \begin{array}{@{}l@{\,}c@{\,}l@{}}
        \varrho^Np-q+
        t_1[a_{i_1},\dots,a_{i_{\mu}}]
        &=&
        DK+R + t_1[a_{i_1},\dots,a_{i_{\mu}}]
        =
        DK +DK'
        \\
        &=& 
        D(K+K')
        \,,
      \end{array}
    \end{equation*}
    and thus, $D\divides \varrho^Np-q+t_1[a_{i_1},\dots,a_{i_{\mu}}]$\@.
    
    First, assume the existence of $a_{i_1},\dots,a_{i_{\mu}}\geq 0$
    with $D\divides R+t_1[a_{i_1},\dots,a_{i_{\mu}}]$, where
    $a_{i_{\xi}}\geq D$, for some $1\leq\xi\leq\mu$\@. To simplify
    matters, we assume without loss of generality that $\xi=1$\@.
    There is an $a\geq 0$ with $a_{i_1}=D+a$\@.
    Further, assume that there is no $b<a_{i_1}$ with $D\divides
    R+t_1[b,a_{i_2},\dots,a_{i_{\mu}}]$\@.  For some $K'\geq 0$, we
    have that 
    \begin{equation*}
      DK'
      \!=\!
      R + t_1[a_{i_1},\dots,a_{i_{\mu}}]
      \!=\!
      R+Dk_{i_1}\!+ t_1[a,a_{i_2},\dots,a_{i_{\mu}}]
      \,.
    \end{equation*}
    Therefore, $D(K'-k_{i_1}) = R + t_1[a,a_{i_2},\dots,a_{i_{\mu}}]$,
    \ie, $D\divides R + t_1[a, a_{i_2},\dots,a_{i_{\mu}}]$\@.  This
    contradicts the minimality of $D+a$\@.
    
    It remains to show the existence of $a_{i_1},\dots,a_{i_{\mu}}\geq
    0$ with $D\divides R+t_1[a_{i_1},\dots,a_{i_{\mu}}]$\@.  The
    existence reduces to the problem of whether the equation
    \begin{equation*}
      D\CDOT y-k_{i_1}\CDOT x_{i_1}-\dots-k_{i_{\mu}}\CDOT x_{i_{\mu}}
      =
      R
    \end{equation*}
    has a solution in the natural numbers. This is the case since
    $\gcd(D,k_{i_1},\dots,k_{i_{\mu}})=1$, by assumption.

    \vspace{\topsep}
    \noindent    
    \textbf{Proof of Step~2:} 
    Assume that there are $\gamma\geq 1$ distinct coefficients in
    $t_2$ of equation~(\ref{eqn:case_++})\@.  Without loss of
    generality, assume that $0<|k_{j_1}|<\dots<|k_{j_{\gamma}}|$\@.
    Let $W:=\frac{\varrho^Np-q+t_1[a_{i_1},\dots,a_{i_{\mu}}]}{D}$
     and
    $\ell_{\xi}:=\frac{|k_{j_{\xi}}|}{D}$, for $1\leq\xi\leq\nu$\@.
    Note that $\ell_1<\dots<\ell_{\gamma}$ and that
    $\gcd(\ell_1,\dots,\ell_{\gamma})=1$\@.
    Equation~(\ref{eqn:case_++}) simplifies with the $a_i$s from
    Step~$1$ to
    \begin{equation}
      \label{eqn:case_++_simplified}
      W
      =
      \ell_1\cdot x_{j_1}+\dots+\ell_{\nu}\cdot x_{j_{\nu}}    
      \,.
    \end{equation}
    An upper bound on $W$ is
    \begin{equation}
      \label{eqn:upper_bound_W}
      \begin{array}{@{}l@{\,}c@{\,}l@{}}
        W
        &\leq&
        \frac{\varrho^Np-q+(\varrho-1)\normpos{t}}{D}
        \leq
        \frac{\varrho^N(\normneg{t}-1)+(\varrho-1)\normpos{t}}{D}
        \\
        &=&
        \frac{\varrho^N\normneg{t}}{D}-
        \frac{\varrho^N}{D}+
        \frac{(\varrho-1)\normpos{t}}{D}
      \end{array}
    \end{equation}
    and a lower bound on $W$ is
    \begin{equation*}
      \begin{array}{@{}l@{\,}c@{\,}l@{}}
        W
        &\geq&
        \frac{\varrho^N-q}{D}
        \geq 
        \frac{\varrho^N-\max\{c,\normneg{t}\}}{D}
        \geq
        \frac{\varrho^{D(\ell_1+\dots+\ell_{\nu})}-D(\ell_1+\dots+\ell_{\nu})}{D}
        \\[3pt]
        &\geq&
        \varrho^{\ell_1+\dots+\ell_{\gamma}}-(\ell_1+\dots+\ell_{\gamma})
        \,.
      \end{array}
    \end{equation*}
    
    From the lower bound on $W$ and the upper bound on Frobenius
    numbers~(\ref{eqn:Frobenius}), it follows that
    equation~(\ref{eqn:case_++_simplified}) has a solution in the
    natural numbers.  Let $\kappa\geq 0$ be maximal such that there
    are $a_1,\dots,a_{\gamma}\geq 0$ with
    \begin{equation}
      \label{eqn:solution_W}
      W
      =
      \ell_1a_1+\dots+\ell_{\gamma}a_{\gamma}+\kappa L
      \,,
    \end{equation}
    where $L:=\frac{\normneg{t}}{D}$\@.  By contradiction, we obtain
    that $a_1,\dots,a_{\gamma}<L$: Assume that there is a $\xi$,
    $1\leq\xi\leq\gamma$ with $a_{\xi}=L+a$, for some $a\geq 0$\@.
    Without loss of generality, assume that $\xi=1$\@.  This
    contradicts the assumption that $\kappa$ is maximal:
    \begin{equation*}
      \begin{array}{@{}l@{\,}c@{\,}l@{}}
        W
        &=&
        \kappa L + \ell_1(L+a) + \ell_2a_2+\dots+\ell_{\gamma}a_{\gamma}
        \\
        &=&
        (\kappa+\ell_1)L + \ell_1a + \ell_2a_2+\dots+\ell_{\gamma}a_{\gamma}
        \,.
      \end{array}
    \end{equation*}
    
    From $\kappa$ and $a_1,\dots,a_{\gamma}$, we obtain a solution for
    equation~(\ref{eqn:case_++_simplified}) in the natural numbers,
    namely
    \begin{equation*}
      \begin{array}{@{}l@{\,}c@{\,}l@{}}
        W
        &=&
        \kappa L+\ell_1a_1+\dots+\ell_{\gamma}a_{\gamma}
        \\
        &=& 
        \kappa (\ell_1+\dots+\ell_{\nu}) +
        \ell_1a_1+\dots+\ell_{\gamma}a_{\gamma}
        \\
        &=&
        \ell_1(\kappa+a_1)+\dots+\ell_{\gamma}(\kappa+a_{\gamma})+
        \ell_{\gamma+1}\kappa+\dots+\ell_{\nu}\kappa
        \,.
      \end{array}
    \end{equation*}
    
    It suffices to show that $\kappa<
    \varrho^N-\max\{a_1,\dots,a_{\gamma}\}$\@.  An upper bound on
    $\kappa$ is
    \begin{equation*}
      \begin{array}{@{}l@{\,}c@{\,}l@{}}
        \kappa
        &\overset{(\ref{eqn:solution_W})}{=}&
        \frac{W-(\ell_1a_1+\dots+\ell_{\gamma}a_{\gamma})}{L}
        \\
        &\leq& 
        \frac{W}{L}-\frac{\max\{a_1,\dots,a_{\gamma}\}}{L} 
        \\
        &\overset{(\ref{eqn:upper_bound_W})}{\leq}&
        \frac{\varrho^N\normneg{t}}{DL}-
        \frac{\varrho^N}{DL}+
        \frac{(\varrho-1)\normpos{t}}{DL}-
        \frac{\max\{a_1,\dots,a_{\gamma}\}}{L}
        \\[3pt]
        &\leq&
        \varrho^N-\frac{\varrho^N}{DL}+
        \frac{(\varrho-1)\normpos{t}-\max\{a_1,\dots,a_{\gamma}\}}{L}
        \,.
      \end{array}
    \end{equation*}
    It remains to check whether the inequality
    \begin{equation*}
      \begin{array}{@{}l@{}}
        \varrho^N-\frac{\varrho^N}{DL}+\frac{(\varrho-1)\normpos{t}-\max\{a_1,\dots,a_{\gamma}\}}{L}
        <
        \varrho^N-\max\{a_1,\dots,a_{\gamma}\}
      \end{array}
    \end{equation*}
    is valid.  The previous inequality simplifies to
    \begin{equation*}
      \begin{array}{@{}c@{}}
        \frac{(\varrho-1)\normpos{t}+\max\{a_1,\dots,a_{\gamma}\}(L-1)}{L}
        <
        \frac{\varrho^N}{DL}
        \,.
      \end{array}
    \end{equation*}
    Multiplying with the common denominator $DL$, the inequality
    simplifies further to
    \begin{equation*}
      D(\varrho-1)\normpos{t}+D\max\{a_1,\dots,a_{\gamma}\}(L-1)
      <
      \varrho^N
      \,.
    \end{equation*}
    Since $\max\{a_1,\dots,a_{\gamma}\}\leq L-1$ and
    $N\geq\normneg{t}+\normpos{t}=DL+\normpos{t}$, it suffices to show
    the validity of the inequality
    \begin{equation}
      \label{eqn:case_++_final_inequality}
      D(\varrho-1)\normpos{t}+D(L-1)^2
      <
      \varrho^{DL+\normpos{t}}
      \,.
    \end{equation}
    It is straightforward to show that the
    inequality~(\ref{eqn:case_++_final_inequality}) is true for all
    $D,L\geq 1$ and $\normpos{t}\geq 0$\@.
  \end{case2}

  \begin{case2}[{\rm\!\!: $p<0$ and $q\leq 0$}]
%
    It suffices to prove that there is a solution
    $a_1,\dots,a_r\in[\varrho^N]$  for the equation
    \begin{equation*}
      \label{eqn:case_--}
      t_1(x_{i_1},\dots,x_{i_{\mu}})
      = 
      \varrho^N|p|-|q|+
      t_2(x_{j_1},\dots,x_{j_{\nu}})
      \,,
    \end{equation*}
    where $t_1$ and $t_2$ are defined as in Case~2\@. This equation is
    similar to equation~(\ref{eqn:case_++}) except $t_1$ and $t_2$ are
    swapped.  We can use a similar argumentation as in Case~2 for showing
    the existence of $a_1,\dots,a_r\in[\varrho^N]$\@.
  \end{case2}

  \begin{case2}[{\rm\!\!: $p>0$ and $q<0$}]
    This case can be solved with Case~1 and Case~2\@. Since $p>0$ and
    $q<0$, we have that $0\in S$\@. By Case~2, the state $0$ is
    reachable from $p$, and by Case~1, $q$ is reachable from state
    $0$\@.
  \end{case2}

  \begin{case2}[{\rm\!\!: $p<0$ and $q>0$}]
    Analogously, this case can be solved by Case~3 and Case~1\@.
    \qed
  \end{case2}
\end{proof}

With Lemma~\ref{lem:S} at hand, it is straightforward to prove for
$\autA^{t\ZAECKZAECK c}_{(m,n)}$ that $p\sim q$ iff $p=q$, for all
$p,q\in S$\@. Therefore, we have that the minimal automaton
representing $\Zrepresents{t\EQ c}$ has at least $|S|$ states.  

Another consequence of Lemma~\ref{lem:S} is that $S$ is a strongly
connected component in $\autA^{t\ZAECKZAECK c}_{(m,n)}$: By
Lemma~\ref{lem:S}, every state $q\in S$ is reachable from every $p\in
S$, and it is easy to show that the initial state $q_{\rmI}$ is not
reachable from a state in $S$ and that a state in $S$ cannot be
reached from any state that is not in $S\cup\{q_{\rmI}\}$\@.

\subsection{Divisibility Relation}
\label{subsec:divisibility}

In this subsection, we give an upper bound of the size of the minimal
DWA for a formula $d\divides t+c$, where $d\geq 2$, $t(x_1,\dots,x_r)$
is a homogeneous term, and $c\in\Int$\@.

Let $\autA^{d\divides t+c}$ be the DWA with the set of states
$Q:=\{q_{\rmI},0,1,\dots,d-1\}$\@. A state $q\in Q\cap\Int$ has an
intuitive interpretation: if we reach the state $q$ with a word $w\in
(\Sigma^r)^*$ then the remainder of the division of $t[\toints{w}]$ by
$d$ equals $q$\@.  We denote by $\remainder(q,d)$ the remainder of
$q\in\Int$ divided by $d$\@.  Let $\autA^{d\divides
  t+c}:=(Q,\Sigma^r,\delta,q_{\rmI},F)$ with
\begin{equation*}
  \delta(q,\overline{b}):=
  \begin{cases}
    \remainder\bigl(t[\sigma(\overline{b})],d\bigr) & 
    \text{if $q=q_{\rmI}$,} \\
    \remainder\bigl(\varrho q+t[\overline{b}],d\bigr)& 
    \text{otherwise,}
  \end{cases}
\end{equation*}
for $q\in Q$ and $\overline{b}\in\Sigma^r$, and $F:=\set{q\in
  Q\cap\Int}{d\divides q+c}$\@. Note that there is exactly one $q\in
Q\cap\Int$ with $d\divides q+c$\@.

The correctness of our construction follows from two facts:
\begin{enumerate}[(a)]
\item \label{enum:divides_remainder}%
  For $n\in\Int$, $d\divides n+c$ iff
  $d\divides\remainder(n,d)+c$\@.
\item \label{enum:transition_function_remainder}%
  For $w\in(\Sigma^r)^+$,
  $\widehat{\delta}(q_{\rmI},w)=\remainder\bigl(t[\toints{w}],d\bigr)$\@.
\end{enumerate}
The proof of~(\ref{enum:divides_remainder}) is straightforward.  There
are $p,q\in\Int$ such that $pd+q=n$ and $0\leq q<d$\@.  Note that
$q=\remainder(n,d)$\@.  By definition, $d\divides n+c$ iff there is a
$k\in\Int$ with $dk=n+c=pd+q+c$\@. The equality can be rewritten to
$d(k-p)=q+c$, \ie, $d\divides \remainder(n,d)+c$\@.
  
We prove~(\ref{enum:transition_function_remainder}) by induction over
the length of $w$\@.  For the base case, let
$w=\overline{b}\in\Sigma^r$\@.  Since we represent integers using
$\varrho$'s complement, we have that
$t[\toints{\overline{b}}]=t[\sigma(\overline{b})]$\@.  By definition,
$\widehat{\delta}(q_{\rmI},\overline{b})=
\remainder\bigl(t[\toints{\overline{b}}],d\bigr)$\@.
For the step case, assume
$\widehat{\delta}(q_{\rmI},w)=\remainder\bigl(t[\toints{w}],d\bigr)$
and let $\overline{b}\in\Sigma^r$\@.
There are $p,q\in\Int$ with $t[\toints{w}]=pd+q$ and $0\leq q<d$\@.
Note that $q=\remainder\bigl(t[\toints{w}],d\bigr)$ and
$t[\toints{w\overline{b}}]=\varrho
t[\toints{w}]+t[\overline{b}]=\varrho pd+\varrho q+t[\overline{b}]$\@.
We have that
\begin{equation*}
  \begin{array}{@{}l@{\,}c@{\,}l@{}}
    \remainder\bigl(t[\toints{w\overline{b}}],d\bigr)
    &=&
    \remainder(\varrho pd+\varrho q+t[\overline{b}],d)
    \\
    &=&
    \remainder(\varrho q+t[\overline{b}],d)=
    \delta(q,\overline{b})
    \\
    &\overset{\textup{IH}}{=}&
    \delta(\widehat{\delta}(q_{\rmI},w),\overline{b})
    =\widehat{\delta}(q_{\rmI},w\overline{b})
    \,.
  \end{array}
\end{equation*}
\begin{lemma}
  \label{fact:bound_automata_divisibiliy}
  The DWA $\autA^{d\divides t+c}$ represents $\Zrepresents{d\divides
    t+c}$ and has $d+1$ states.
\end{lemma}

An optimization of the construction is to filter out the states that
are not a multiple of $\gcd(\gcd(t),d)$\@. These states are not
reachable from the initial state since $\remainder(t[\overline{a}],d)$
is a multiple of $\gcd(\gcd(t),d)$, for every
$\overline{a}\in\Int^r$\@. 

\subsection{Quantifier-free Formulas}
\label{subsec:qf_formulas}

In this subsection, we give an upper bound on the size of the minimal
DWA for a quantifier-free PA formula.  This upper bound depends on the
maximal absolute value of the constants occurring in the
(in)equa\-tions of the formula, the homogeneous terms, and the
divisibility relations.  The upper bound does \emph{not} depend on the
Boolean combination of the atomic formulas.
This is not obvious since Boolean connectives are handled by the
product construction if we construct the DWA recursively over the
structure of the quantifier-free formula. The size of the resultant
DWA using the product construction is in the worst case the product of
the number of states of the two DWAs.
%

Let $\sfT$ be a finite nonempty set of homogeneous terms and let
$\sfD$ be a finite set of atomic formulas of the form $d\divides t$,
where $d\geq 1$ and $t$ is a homogeneous term.  Moreover, let
$\ell>\max\set{\normpos{t}}{t\in \sfT}\cup\set{\normneg{t}}{t\in
  \sfT}$ and $\ell'>\max\set{d}{d\divides t\in\sfD}$\@.
\begin{theorem}
  \label{thm:bound_qf_formula}
  Let $\psi$ be a Boolean combination of atomic formulas $t\ZAECKZAECK
  c$ and $d\divides t+c'$, with $t\in\sfT$, $d\divides t\in\sfD$,
  $-\ell<c<\ell$, $c'\in\Int$, and
  $\zaeckzaeck\in\{=,\not=,<,\leq,>,\geq\}$\@. The size of the minimal
  DWA for $\psi$ is at most $(2+2\ell)^{|\sfT|}\cdot
  \ell'^{|\sfD|}$\@.
\end{theorem}
\begin{proof}
  Without loss of generality, we assume that the variables occurring
  in terms in $T$ are $y_1,\dots,y_r$\@.
  Let $\autC$ be the product automaton of all the
  $\autA_{(-\ell,\ell)}^{t\EQ 0}$s and $\autA^{d\divides t}$s, for
  $t\in\sfT$ and $d\divides t\in\sfD$\@.  To simplify notation we omit
  the subscripts $(-\ell,\ell)$ and we assume that
  $\sfT=\{t_1,\dots,t_m\}$ and $\sfD=\{d_1\divides
  t_1,\dots,d_n\divides t_n\}$\@.
  Note that the states of $\autC$ are tuples
  $(p_1,\dots,p_m,q_1,\dots,q_n)$, where $p_i$ is a state of
  $\autA^{t_i\EQ 0}$ and $q_j$ is a state of $\autA^{d_j\divides
    t_j}$\@. By Lemma~\ref{fact:bound_automata_(in)equation},
  $\autA^{t_i\EQ 0}$ has $2+2\ell$ states, and by
  Lemma~\ref{fact:bound_automata_divisibiliy}, $\autA^{d_j\divides
    t_j}$ has $1+d_j\leq \ell'$ states.
  %
  %
  It follows that the size of $\autC$ is at most
  \begin{equation*}
      \prod_{t\in\sf T}(2+2\ell)\cdot
      \prod_{d\divides t\in\sfD} (1+d)
      \leq
      (2+2\ell)^{|\sfT|}\cdot
      \ell'^{|\sfD|}
      \,.
  \end{equation*}
  
  It remains to define the set of accepting states of $\autC$
  according to $\psi$\@.
  %
  We define the DWA $\autD$ as $\autC$ except the set $E$ of accepting
  states is defined as follows.  A state
  $q=(p_1,\dots,p_m,q_1,\dots,q_n)\in\Int^{m+n}$
  of $\autD$ is in $E$ iff $\frakZ\models\psi_q$, where $\psi_q$ is
  the formula obtained by substituting
  \begin{itemize}
  \item the integer $p_i$ for the term $t_i$ in the atomic
    formulas of the form $t_i\ZAECKZAECK c$, and
  \item the integer $q_j$ for the term $t_j$ in the atomic formulas of
    the form $d_j\divides t_j+c$\@.
  \end{itemize}
  Note that $\psi_q$ is either true or not in $\frakZ$ since it is a
  sentence.
 
  It remains to prove that $\autD$ represents $\Zrepresents{\psi}$\@.
  Let $w\in(\Sigma^r)^+$ be a word representing
  $\overline{a}\in\Int^r$\@.
  For a term $t\in\sfT$, the value $t[\overline{a}]$ can be replaced
  by $\ell$ if $t[\overline{a}]\geq \ell$ and by $-\ell$ if
  $t[\overline{a}]\leq -\ell$ in every atomic formula of the form
  $t\ZAECKZAECK c$ without changing its truth value since
  $-\ell<c<\ell$\@.  This modified value corresponds to the state
  reached by $\autA^{t\EQ 0}$ after reading the word $w$\@.  
  For an atomic formula of the form $d\divides t+c$, with $d\divides
  t\in\sfD$, we can replace $t[\overline{a}]+c$ by
  $\remainder(t[\overline{a}]+c,d)$ without changing the truth value.
  This adjusted value corresponds to the state reached by
  $\autA^{d\divides t}$ after reading the word $w$\@.
  From the definition of $E$, it follows that $w\in L(\autD)$ iff
  $\frakZ\models\psi[\overline{a}]$\@.
\end{proof}

\section{An Upper Bound on the Automata Size}
\label{BOUND}

In this section, we give an upper bound on the size of the minimal DWA
for PA formulas.  We obtain this bound by examining the
quantifier-free formulas constructed by applying Reddy and Loveland's
quantifier elimination method~\cite{Reddy_Loveland.1978}, which
improves Cooper's quantifier elimination method~\cite{Cooper.1972}\@.
We use Reddy and Loveland's quantifier elimination method since the
produced formulas are ``small'' with respect to the following
parameters on which the upper bound of the minimal DWA in
Theorem~\ref{thm:bound_qf_formula} depends.
\begin{definition}
  For $\varphi\in\pa$, we define
  \begin{align*}
    \Terms(\varphi)&:=
    \set{t}{t\ZAECKZAECK c\in\Atoms(\varphi)}
    \,,
    \\
    \Divs(\varphi)&:=
    \set{d\divides t}{d\divides t+c\in\Atoms(\varphi)}
    \,,
    \\
    \intertext{and}
    \maxCoef(\varphi)&:=
    \max\{1\}\cup
    \set{\abs{k}}{\text{$k$ is a } \text{coefficient
        in $t\ZAECKZAECK c\in\Atoms(\varphi)$}}
    \,,
    \\
    \maxConst(\varphi)&:=
    \max\{1\}\cup
    \set{\abs{c}}{t\ZAECKZAECK c\in\Atoms(\varphi)}
    \,,
    \\
    \maxDiv(\varphi)&:=
    \max\{1\}\cup\set{d}{d\divides t+c\in\Atoms(\varphi)}
    \,.
  \end{align*}
\end{definition}

\subsection{Eliminating a Quantifier}

For the sake of completeness, we briefly recall Reddy and Loveland's
quantifier elimination method. Consider the formula $\exists x\varphi$
with $\varphi(x,\overline{y})\in\qf$\@.
The construction of $\psi(\overline{y})\in\qf$ proceeds in 2 steps.

\vspace{\topsep}
\noindent
\textbf{Step 1:} First, eliminate the connectives $\rightarrow$ and
$\leftrightarrow$ in $\varphi$ using standard rules, \eg, a subformula
$\chi\IMPL\chi'$ is replaced by $\NEG\chi\OR\chi'$\@.
Second, push all negation symbols in $\varphi$ inward (using
De\,Morgan's laws, etc.\@) until they only occur directly in front
of the atomic formulas.
Third, rewrite all atomic formulas and negated atomic formulas in
which $x$ occurs such that they are all of one of the forms
\begin{gather}
  \tag{A}
  \label{eqn:atom_typeA}
  k\CDOT x\LESS t(y_1,\dots,y_n)
  \,,
  \\
  \tag{B}
  \label{eqn:atom_typeB}
  t(y_1,\dots,y_n)\LESS k\CDOT x
  \,,
  \\
  \intertext{or}
  \tag{C}
  \label{eqn:atom_typeC}
  d\DIVIDES t(x,y_1,\dots,y_n)
\end{gather}
with $k>0$\@. 
For instance, the negated inequation $\NEG 2\CDOT x+9\CDOT y\LESS 5$
is rewritten to $-9\CDOT y+5-1\LESS 2\CDOT x$, and the negated
equation $\NEG 2\CDOT x+9\CDOT y\EQ 5$ is replaced by the disjunction
$-9\CDOT y+5\LESS 2\CDOT x\OR 2\CDOT x\LESS -9\CDOT y+5$\@.
Let $\varphi'(x,\overline{y})$ be the resulting formula.

\vspace{\topsep}
\noindent
\textbf{Step 2:} Let $\psi_{-\infty}$ be the formula where all the
atomic formulas of type~(\ref{eqn:atom_typeA}) in $\varphi'$ are
replaced by ``true'', \ie, $0\LESS1$, and all atomic formulas of
type~(\ref{eqn:atom_typeB}) are replaced by ``false'', \ie,
$1\LESS0$\@.  We assume in the following, without loss of generality,
that $0\LESS1$ and $1\LESS0$ do not occur as proper subformulas.  Note
that by propositional reasoning, we can always eliminate such
subformulas, \eg, $\alpha\AND 0\LESS1$ can be simplified to $\alpha$\@.
Let $\sfB$ be the set of the atomic formulas in $\varphi'$ of
type~(\ref{eqn:atom_typeB}), and let $\lcm(x,\varphi)$ be the least
common multiple of the $d$s in the atomic formulas of
type~(\ref{eqn:atom_typeC}) and of the coefficients of the variable
$x$ in the atomic formulas of type~(\ref{eqn:atom_typeB})\@.  Let
$\psi$ be the formula
\begin{equation*}
  \bigvee_{1\leq j\leq \lcm(x,\varphi)}\psi_{-\infty}[j/x]
  \,\,
  \OR
  \bigvee_{1\leq j\leq \lcm(x,\varphi)\ }
  \bigvee_{t+c\LESS k\cdot x\in\sfB}
  \bigl(
  k\!\DIVIDES\! t+c+j\AND\varphi'[t+c+j/k\cdot x]
  \bigr)
  \,,
\end{equation*}
where $\varphi'[t+c+j/k\cdot x]$ means that every atomic formula
$\alpha$ in $\varphi'$ in which $x$ occurs is first multiplied by $k$
and then $k\cdot x$ is substituted by $t+c+j$\@. Formally, for an
atomic formula $\alpha$, a term $t$, and $k\in\Int\setminus\{0\}$, we
define
\begin{equation*}
  \alpha[t/k\cdot x]:=
  \begin{cases}
    k'\cdot t \LESS k\cdot t' &
    \text{if $\alpha=k'\cdot x\LESS t'$,} 
    \\
    k\cdot t' \LESS k'\cdot t &
    \text{if $\alpha=t'\LESS k'\cdot x$,} 
    \\
    kd\divides k'\cdot t+k\cdot t' &
    \text{if $\alpha=d\divides k'\cdot x + t'$,}
    \\
    \alpha & \text{otherwise.}
  \end{cases}
\end{equation*}


\begin{fact}
  The formula $\psi$ is logically equivalent to $\exists x\varphi$\@.
\end{fact}

\subsection{Analysis}

We can construct from an arbitrary formula a logically equivalent
quantifier-free formula by successively replacing subformulas of the
form $Qx\varphi$, where $\varphi\in\qf$ and $Q\in\{\exists,\forall\}$,
with the logically equivalent quantifier-free formulas that are
produced by the quantifier elimination method.
\citeN{Oppen.1978} analyzed the length of the formulas that are
produced by iteratively applying Cooper's quantifier elimination
method. Oppen proved a triple exponential upper bound on the formula
length by relating the growth in the number of atomic formulas, the
maximum of the absolute values of constants and coefficients appearing
in these atomic formulas, and the number of distinct coefficients and
divisibility predicates that may appear.  
Similar analysis of improved versions of Cooper's quantifier
elimination method are in~\cite{Reddy_Loveland.1978,Graedel.1988}\@.

\citeN{Reddy_Loveland.1978} observed that they obtain shorter formulas
when pushing quantifiers inward before applying their quantifier
elimination method.  For example, using the quantifier elimination
method to eliminate the quantified variable $x_2$ in $\exists
x_1\exists x_2\varphi$ with $\varphi\in\qf$, we obtain a formula of
the form $\exists x_1(\varphi_1\OR\dots\OR\varphi_n)$\@.  Instead of
applying the quantifier elimination method to $\exists
x_1(\varphi_1\OR\dots\OR\varphi_n)$, rewriting the formula first to
$(\exists x_1\varphi_1)\OR\dots\OR(\exists x_1\varphi_n)$ and then
applying the quantifier elimination method to each of the disjuncts
separately produces shorter formulas due to the following reasons.
First, we avoid using $\lcm(x_1,\varphi_1\OR\dots\OR\varphi_n)$ in
Step~2 of the quantifier elimination method; instead we determine
$\lcm(x_1,\varphi_i)$, for each disjunct $\varphi_i$ separately.
Second, we use an inequation $k\cdot x_1\LESS t$ of
type~(\ref{eqn:atom_typeB}) occurring in a disjunct $\varphi_i$ only
for eliminating $x_1$ in $\varphi_i$\@. We do not use this inequation
$k\cdot x_1\LESS t$ for eliminating $x_1$ in disjuncts $\varphi_j$ in
which the inequation $k\cdot x_1\LESS t$ does not occur.
However, if the variable $x_1$ is universally quantified, then we
cannot push the quantifier inward. Note that in order to apply the
quantifier elimination method, we have to rewrite the formula $\forall
x_1(\varphi_1\OR\dots\OR\varphi_n)$ to $\neg\exists
x_1(\neg(\varphi_1\OR\dots\OR\varphi_n))$\@.  To eliminate $x_1$, we
have to use in Step~2 $\lcm(x_1,\neg(\varphi_1\OR\dots\OR\varphi_n))$
and the set $\sfB$ of the inequations of type~(\ref{eqn:atom_typeB})
occurring in the formula produced by Step~1 normalizing
$\neg(\varphi_1\OR\dots\OR\varphi_n)$\@.

Reddy and Loveland analyzed the quantifier-free formulas produced by
successively applying their quantifier elimination method to formulas
in prenex normal form.  We refine and extend their analysis to
arbitrary formulas. 
However, before launching into the analysis, we need the following
definitions.  For $\varphi\in\pa$, we define
\begin{align*}
  \Terms_+(\varphi)&:= 
  \set{t\in\Terms(\varphi)}
  {\text{in $t$ there occurs a variable that is bound in $\varphi$}} 
  \\
  \intertext{and}
  \Divs_+(\varphi)&:= 
  \set{d\divides t\in\Divs(\varphi)}
  {\text{in $t$ there occurs a variable that is bound in $\varphi$}} 
  \,.
\end{align*}
Furthermore, let
$\Terms_-(\varphi):=\Terms(\varphi)\setminus\Terms_+(\varphi)$ and
$\Divs_-(\varphi):=\Divs(\varphi)\setminus\Divs_+(\varphi)$\@.

\begin{lemma}
  \label{lem:qe_bounds}
  For every $\varphi\in\pa$ of the form $Q x_1\dots Q x_s\vartheta$,
  with $Q\in\{\exists,\forall\}$ and $\vartheta\in\qf$, there is a
  logically equivalent $\psi\in\qf$ such that
  \begin{align*}
    |\Terms(\psi)\setminus\Terms_-(\varphi)|
    &\leq 
    |\Terms_+(\varphi)|^{s+1}
    \,,\\
    |\Divs(\psi)\setminus\Divs_-(\varphi)| 
    &\leq 
    \big(|\Terms_+(\varphi)|+1\big)^s\cdot
    \big(|\Divs_+(\varphi)|+s\big)
    \,,\\
    \intertext{and}
    \maxCoef(\psi) 
    &< 
    a^{2^{2s}}
    \,,\\
    \maxDiv(\psi) 
    &< 
    a^{2^{2s}}
    \,,\\
    \maxConst(\psi) 
    &<  
    ba^{2^{2s}
      (|\Terms_+(\varphi)|+|\Divs_+(\varphi)|+s)}
    \,,
  \end{align*}
  where $a>\max\{2,\maxCoef(\varphi),\maxDiv(\varphi)\}$ and
  $b>\max\{2,\maxConst(\varphi)\}$\@.
\end{lemma}
\begin{proof}
  We first describe how we construct the quantifier-free formula
  $\psi$, where we assume that $Q=\exists$\@.  For $Q=\forall$, we
  rewrite $\varphi$ to $\neg\exists x_1\dots\exists x_s\neg\vartheta$
  and eliminate the quantified variables in $\exists x_1\dots\exists
  x_s\neg\vartheta$ as described below.
  
  By a preprocessing step we rewrite $\vartheta$ to negation norm form
  (\ie, we eliminate the connectives $\rightarrow$ and
  $\leftrightarrow$, and we push the negation symbols inward such that
  the connective $\neg$ only occurs directly in front of atomic
  formulas) and we rewrite (in)equations so that we only have
  inequations of the form $t\LESS t'$ or $t\GREATER t'$ and no
  negation occurs in front of an inequation. For instance, $t\LEQ t'$
  is rewritten to $t\LESS t'+1$ and $\neg t\LEQ t'$ is rewritten to
  $t\GREATER t'$\@.  Let $\vartheta_0$ be the formula that we obtain
  by the rewriting.
  The only parameter that is changed by this rewriting is the maximal
  absolute value of a constant, which increases by at most $1$\@.
  Observe that this special form of a formula is preserved when we
  apply the quantifier elimination method: In Step~1 we only rewrite
  the inequations such that they are of type~(\ref{eqn:atom_typeA})
  or~(\ref{eqn:atom_typeB})\@.  Such rewriting does not alter the
  parameters.  Step~2 also preserves this special form.
  
  After the preprocessing step, we construct the quantifier-free
  formula $\psi$ iteratively in $s$ steps by constructing intermediate
  formulas $\varphi_0,\dots,\varphi_s$, where $\psi$ will be
  $\varphi_s$. Let $\varphi_0:=\exists x_1\dots\exists
  x_s\vartheta_0$\@.
  In the $\ell$th step we eliminate the variable $x_{s-\ell+1}$, where
  $1\leq \ell\leq s$\@. This is done as follows.  Assume that
  $\varphi_{\ell-1}=\exists x_1\dots\exists
  x_{s-\ell+1}\vartheta_{\ell-1}$, where $\vartheta_{\ell-1}=
  \vartheta_{\ell-1,1}\OR\dots\OR\vartheta_{\ell-1,n_{\ell-1}}$\@.
  We push the existential quantification of $x_{s-\ell+1}$ inward in
  $\vartheta_{\ell-1}$ as far as possible\@.  For every $1\leq i\leq
  n_{\ell-1}$, we apply the quantifier elimination method to $\exists
  x_{s-\ell+1}\vartheta_{\ell-1,i}$\@.  After the $n_{\ell-1}$
  applications of the quantifier elimination method, we obtain for
  some $n_\ell\geq 1$, a formula
  $\vartheta_\ell:=\vartheta_{\ell,1}\OR\dots\OR\vartheta_{\ell,n_\ell}$
  that is logically equivalent to $\exists
  x_{s-\ell_1}\vartheta_{\ell-1}$\@.  Let $\varphi_\ell:=\exists
  x_1\dots\exists x_{s-\ell}\vartheta_\ell$\@.

  \medskip 
  
  We now prove the upper bounds on the parameters of $\psi$\@.  Let
  $n_0:=1$ and $\vartheta_{0,1}:=\vartheta_0$\@. It is straightforward
  to prove by induction over $0\leq \ell\leq s$:
  \begin{enumerate}[(i)]
  \item There are indices $1\leq i_1,\dots,i_k\leq n_\ell$ such that
    \begin{equation*}
      \Terms(\varphi_\ell) =
      \Terms(\vartheta_{\ell,i_1})
      \cup\dots\cup
      \Terms(\vartheta_{\ell,i_k})
      \,,
    \end{equation*}
    where $k\leq |\Terms_+(\varphi)|^\ell$\@.
  \item There are indices $1\leq i_1,\dots,i_k\leq n_\ell$ such that
    \begin{equation*}
      \Divs(\varphi_\ell) =
      \Divs(\vartheta_{\ell,i_1})
      \cup\dots\cup
      \Divs(\vartheta_{\ell,i_k})
      \,,
    \end{equation*}
    where $k\leq (|\Terms_+(\varphi)|+1)^\ell$\@. 
  \end{enumerate}  
  The upper bounds on $|\Terms(\psi)\setminus\Terms_-(\varphi)|$ and
  $|\Divs(\psi)\setminus\Divs_-(\varphi)|$ follow immediately from~(i)
  and~(ii), respectively, since $|\Terms(\vartheta_{\ell,i})\setminus
  \Terms_-(\varphi)|\leq|\Terms_+(\varphi)|$ and
  $|\Divs(\vartheta_{\ell,i})\setminus\Divs_-(\varphi)|\leq
  |\Divs_+(\varphi)|+\ell$, for every $0\leq \ell\leq s$ and $1\leq i\leq
  n_\ell$\@.
  
  \medskip
 
  We establish upper bounds on $\maxCoef(\psi)$, $\maxDiv(\psi)$, and
  $\maxConst(\psi)$:
  We prove by induction over $\ell$ that
  \begin{equation*}
    \maxCoef(\varphi_\ell), \maxDiv(\varphi_\ell)
    <
    a^{2^{2\ell}}
    \quad\text{and}\quad
    \maxConst(\varphi_{\ell})
    < 
    ba^{2^{2\ell}(|\Terms_+(\varphi)|+|\Divs_+(\varphi)|+\ell)}
    \,.
  \end{equation*}
  For $\ell=0$, these upper bounds are obviously true.  Assume that
  $\ell>0$\@.  For $1\leq i\leq n_{\ell-1}$, we examine at the formula
  produced by the quantifier elimination method applied to $\exists
  x_{s-\ell+1}\vartheta_{\ell-1,i}$\@.  Note that Step~1 of the
  quantifier elimination method does not alter the absolute values of
  the coefficients and constants, and the $d$s in the divisibility
  predicate because of our preprocessing step by rewriting $\vartheta$
  to $\vartheta_0$\@.  It suffices to look at the substitutions
  $\alpha[t+c+j/k\cdot x]$ carried out in Step~2, where $\alpha$ is an
  atomic formula in $\vartheta_{\ell-1,i}$, $t+c\LESS k\cdot x$ is an
  inequation of type~(\ref{eqn:atom_typeB}) in $\vartheta_{\ell-1,i}$,
  and $1\leq j\leq\lcm(x_{s-\ell+1},\vartheta_{\ell-1,i})$\@.
  \begin{itemize}
  \item Assume that $\alpha=d\divides t$, for some $d\geq 1$ and some
    term $t$\@.  By the induction hypothesis, we have that
    \begin{equation*}
      kd<a^{2^{2(\ell-1)}}\cdot a^{2^{2(\ell-1)}}
      =
      a^{2\cdot 2^{\ell-1}}
      \leq 
      a^{2^{2\ell}}
      \,.
    \end{equation*}
    It follows that $\maxDiv(\varphi_{\ell})< a^{2^{2\ell}}$\@.
  \item Assume that $\alpha=k'\cdot x\LESS t'$ or $\alpha=t'\LESS
    k'\cdot x$, for some $k'>0$ and some term $t'$\@.  By the induction
    hypothesis, we have that $k$, $k'$, and the absolute values of the
    coefficients occurring in $t$ and $t'$ are smaller than 
    $a^{2^{2(\ell-1)}}$\@.  It follows that 
    the absolute values of the coefficients in the normalized
    inequations of $k'\cdot (t+c+j)\LESS k\cdot t'$ and $k\cdot
    t'\LESS k'\cdot (t+c+j)$ are smaller than
    \begin{equation*}
      a^{2^{2(\ell-1)}}\cdot a^{2^{2(\ell-1)}} +
      a^{2^{2(\ell-1)}}\cdot a^{2^{2(\ell-1)}}
      =
      2a^{2^{2\ell-1}}
      \leq 
      a^{2^{2\ell}}
      \,.
    \end{equation*}
    Hence, $\maxCoef(\varphi_{\ell})<a^{2^{2\ell}}$\@.
  
    The absolute values of the constants in the normalized inequations
    $k'\cdot (t+c+j)\LESS k\cdot t'$ and $k\cdot t'\LESS k'\cdot
    (t+c+j)$ is bounded by
    \begin{multline*}
      \maxCoef(\varphi_{\ell-1})
      \cdot
      \big(\maxConst(\varphi_{\ell-1})+
      \lcm(x_{s-\ell+1},\vartheta_{\ell-1,i})\big)
      +
      \\
      \maxCoef(\varphi_{\ell-1})\cdot\maxConst(\varphi_{\ell-1})
      \,,
    \end{multline*}
    which rewrites to
    \begin{equation}
      \label{eqn:upper_constant}
      \maxCoef(\varphi_{\ell-1})
      \cdot
      \big(2\maxConst(\varphi_{\ell-1})+
      \lcm(x_{s-\ell+1},\vartheta_{\ell-1,i})\big)
      \,.
    \end{equation}
    An upper bound on $\lcm(x_{s-\ell+1},\vartheta_{\ell-1,i})$ is
    \begin{equation*}
      \big(a^{2^{2(\ell-1)}}\big)^{|\Terms_+(\varphi)|+|\Divs_+(\varphi)|+\ell-1}
      =
      a^{2^{2(\ell-1)}\cdot(|\Terms_+(\varphi)|+|\Divs_+(\varphi)|+\ell-1)}
    \end{equation*}
    since we determine the least common multiple of at most
    $|\Terms_+(\varphi)|+|\Divs_+(\varphi)|+\ell-1$ numbers and all
    these numbers are bounded by $a^{2^{2(\ell-1)}}$\@.
    By the induction hypothesis, we have that $|c|$ and the absolute value
    of the constant in $t'$ is smaller than
    $ba^{2^{2(\ell-1)}(|\Terms_+(\varphi)|+|\Divs_+(\varphi)|+\ell-1)}$\@.
    Therefore,~(\ref{eqn:upper_constant}) is smaller than
    \begin{align*}
      a^{2^{2\ell-1}}
      \big(2ba^{|\Terms_+(\varphi)|+|\Divs_+(\varphi)|+\ell-1}+
      a^{|\Terms_+(\varphi)|+|\Divs_+(\varphi)|+\ell-1}\big)
      &\leq
      2ba^{2^{2\ell}(|\Terms_+(\varphi)|+|\Divs_+(\varphi)|+\ell-1)}
      \\
      &\leq
      ba^{2^{2\ell}(|\Terms_+(\varphi)|+|\Divs_+(\varphi)|+\ell)}
      \,.
    \end{align*}
    It follows that $\maxConst(\varphi_{\ell})<
    ba^{2^{2\ell}(|\Terms_+(\varphi)|+|\Divs_+(\varphi)|+\ell)}$\@.\qed
  \end{itemize}    
\end{proof}

By iteratively applying Lemma~\ref{lem:qe_bounds} we obtain the
following upper bounds for formulas in prenex normal form.
\begin{lemma}
  \label{lem:qe*_bounds}
  For every $\varphi\in\pa$ of the form $Q_1 x_1\dots Q_r x_r \psi_0$
  with $\psi_0\in\qf$ there is logically equivalent $\psi\in\qf$ such
  that
  \begin{equation*}
    |\Terms(\psi)| \leq 
    T^{(\ell+1)^{\qa(\varphi)}}
    \qquad\text{and}\qquad
    |\Divs(\psi)| \leq
    DT^{(\ell+1)^{\qa(\varphi)+2}}
    \,,
  \end{equation*}
  where $T=\max\{2,|\Terms(\varphi)|\}$,
  $D=\max\{1,|\Divs(\varphi)|\}$, and $\ell$ is the maximal length of
  a quantifier block in $\varphi$\@. Furthermore, it holds that
  \begin{align*}
    \maxCoef(\psi) 
    &<
    a^{2^{2\qn(\varphi)}}
    \,,\\
    \maxDiv(\psi) 
    &<
    a^{2^{2\qn(\varphi)}}
    \,,\\
    \intertext{and}
    \maxConst(\psi) &<
    ba^{2^{3\qn(\varphi)}DT^{(\ell+1)^{\qa(\varphi)+2}}}
    \,,
  \end{align*}
  where $a>\max\{2,\maxCoef(\varphi),\maxDiv(\varphi)\}$ and
  $b>\max\{2,\maxConst(\varphi)\}$\@.
\end{lemma}
\begin{proof}
  We construct the quantifier-free formula $\psi$ by successively
  eliminating the quantifier blocks in $\varphi$, starting from the
  innermost block.  Assume that after the $k$th step, where $0\leq
  k<\qa(\varphi)$, we have produced the formula
  \begin{equation*}
    Q_1 x_1\dots Q_i x_i Q x_{i+1}\dots Q x_j \psi_k
    \,,
  \end{equation*}
  where $1\leq i<j\leq r$, $Q_1,\dots,Q_i,Q\in\{\exists,\forall\}$
  with $Q_i\not=Q$, and $\psi_k\in\qf$\@.  Let $\psi_{k+1}\in\qf$ be
  the formula from Lemma~\ref{lem:qe_bounds} that is logically
  equivalent to $\varphi_k:=Q x_{i+1}\dots Q x_j \psi_k$\@.  We define
  $\psi:=\psi_{\qa(\varphi)}$\@.
  
  For $1\leq i\leq\qa(\varphi)$, let $\ell_i$ be the length of the
  $i$th quantifier block.  We prove by induction over $0\leq
  k\leq\qa(\varphi)$ that
  \begin{align*}
    |\Terms(\psi_k)|
    \leq 
    T^{(\ell+1)^k}
    \qquad&\text{and}\qquad
    |\Divs(\psi_k)|\leq 
    DT^{(\ell+1)^{k+2}}
    \,,
    \\
    \maxCoef(\psi_k)
    < a^{2^{2(\ell_1+\dots+\ell_k)}}
    \qquad&\text{and}\qquad
    \maxDiv(\psi_k)
    < a^{2^{2(\ell_1+\dots+\ell_k)}}
    \,,
    \\
    \intertext{and}
    \maxConst(\psi_k) 
    &<
    ba^{2^{3(\ell_1+\dots+\ell_k)}
      DT^{(\ell+1)^{k+2}}}
    \,.
  \end{align*}
  The base cases for $k=0$ are trivial.  For the step cases, let
  $k>0$\@.
  \begin{enumerate}[1.]
  \item By Lemma~\ref{lem:qe_bounds}, we have that
    \begin{align*}
      |\Terms(\psi_k)\setminus \Terms_-(\varphi_{k-1})|
      &\leq
      |\Terms_+(\varphi_{k-1})|^{\ell+1}
      \\
      &\leq
      |\Terms(\psi_{k-1})|^{\ell+1}
      \overset{\rm IH}{\leq}
      \big(T^{(\ell+1)^{k-1}}\big)^{\ell+1}
      =
      T^{(\ell+1)^k}
    \end{align*}
    and
    \begin{align*}
      |\Divs(\psi_k)\setminus\Divs_-(\varphi_{k-1})|
      &\leq 
      (|\Terms_+(\varphi_{k-1})|+1)^{\ell}\cdot(|\Divs_+(\varphi_{k-1})|+\ell)
      \\
      &
      \leq
      (|\Terms(\psi_{k-1})|+1)^{\ell}\cdot(|\Divs(\psi_{k-1})|+\ell)
      \\
      &\overset{\rm IH}{\leq}
      \big(T^{(\ell+1)^{k-1}}+1\big)^{\ell}\cdot
      \big(DT^{{(\ell+1)}^{k+1}}+\ell\big)
      \\
      &\leq
      2^{\ell+1}DT^{(\ell+1)^k+(\ell+1)^{k+1}}
      \leq
      DT^{(\ell+1)+(\ell+1)^k+(\ell+1)^{k+1}}
      \\
      &\leq
      DT^{{(\ell+1)}^{k+2}}
      \,.
    \end{align*}
    Note that $T\geq 2$ and $D\geq 1$\@.
    
  \item By Lemma~\ref{lem:qe_bounds}, we have that
    \begin{align*}
      \maxCoef(\psi_k)
      &\leq
      \big(\max\{2,\maxCoef(\psi_{k-1})\}\big)^{2^{2\ell_k}}
      \\
      &\overset{\rm IH}{<}
      \big(a^{2^{2(\ell_1+\dots+\ell_{k-1})}}\big)^{2^{2\ell_k}}
      =
      a^{2^{2(\ell_1+\dots+\ell_k)}}
      \,.
    \end{align*}
    Analogously, we obtain the upper bound for $\maxDiv(\psi_k)$\@.
  
  \item By Lemma~\ref{lem:qe_bounds}, we have that
    \begin{align*}
      \maxConst(\psi_k)
      &\leq
      \maxConst(\psi_{k-1})
      \cdot
      \big(a^{2^{2(\ell_1+\dots+\ell_{k-1})}}\big)^{2^{2\ell_k}
        (|\Terms_+(\varphi_{k-1})|+|\Divs_+(\varphi_{k-1})|+\ell_k)}
      \\
      &\leq
      \maxConst(\psi_{k-1}) 
      a^{2^{2(\ell_1+\dots+\ell_k)}
        (|\Terms(\psi_{k-1})|+|\Divs(\psi_{k-1})|+\ell_k)}
      \\
      &
      \leq
      \maxConst(\psi_{k-1}) 
      a^{2^{2(\ell_1+\dots+\ell_k)}
        (T^{(\ell+1)^{k-1}}+DT^{(\ell+1)^{k+1}}+\ell_k)}
      \\
      &\leq
      \maxConst(\psi_{k-1}) 
      a^{2^{2(\ell_1+\dots+\ell_k)}
        (DT^{(\ell+1)^k}+DT^{(\ell+1)^{k+1}})}
      \\
      &\leq
      \maxConst(\psi_{k-1}) 
      a^{2^{2(\ell_1+\dots+\ell_k)}
        DT^{(\ell+1)^{k+2}}}
      \\
      &\overset{\rm IH}{<}
      ba^{2^{3(\ell_1+\dots+\ell_{k-1})}
        DT^{(\ell+1)^{k+1}}}
      \cdot
      a^{2^{2(\ell_1+\dots+\ell_k)}
        DT^{(\ell+1)^{k+2}}}
      \\
      &\leq
      ba^{(2^{3(\ell_1+\dots+s_{k-1})}+2^{2(\ell_1+\dots+\ell_k)})
        DT^{(\ell+1)^{k+2}}}
      \\
      &\leq
      ba^{2^{3(\ell_1+\dots+\ell_k)}
        DT^{(\ell+1)^{k+2}}}
      \,.
      \qed
    \end{align*}
  \end{enumerate}
\end{proof}

Before we generalize Lemma~\ref{lem:qe*_bounds} to arbitrary formulas,
we want to point out that transforming a formula first into prenex
normal form and then eliminating the quantifiers is not a good thing
to do. The formula size can increase because of the following reasons.

First, a transformation into prenex normal form can increase the
number of quantifier alternations. For instance, any transformation of
$(\forall x\varphi)\wedge(\exists y\psi)$ into prenex normal form will
introduce at least one additional alternation of quantifiers.
 
Second, when transforming a formula into prenex normal form we have to
introduce fresh variables when pushing quantifiers to the front.  As
an example, consider the formula in prenex normal form
\begin{align*}
  \exists z_{n-1}\dots\exists z_2\exists z_1 
  (&x=z_{n-1}+z_{n-1}\AND
  \\&
  z_{n-1}=z_{n-2}+z_{n-2}\AND\dots\AND z_2=z_1+z_1\AND z_1=y+y)
  \,,
\end{align*}
for some $n\geq 1$\@.  It consists of $n$ distinct equations.  A
logically equivalent formula that consists of at most $4$ distinct
equations is
\begin{align*}
  \exists z\big(&
  x=z+z\AND
  \\&
  \exists z'(z=z'+z'
  \AND\dots\AND 
  \exists z'(z=z'+z'\AND\exists z(z'=z+z\AND z=y+y))
  \dots)\big)
  \,.
\end{align*}
Furthermore, the formula length decreases by a factor of $\BigOh(\log
n)$ since we use a fixed number of variables, \ie, we use $x,y,z,z'$
instead of $x,y,z_1,\dots,z_{n-1}$\@.

The third reason why a transformation into prenex normal form is not a
good idea is illustrated by the formula $(\forall
x\varphi)\IFF\psi$\@.  Quantifiers do in general not distribute over
$\rightarrow$ and $\leftrightarrow$\@.  Therefore, we eliminate the
connective $\leftrightarrow$ and obtain $((\forall
x\varphi)\IMPL\psi)\AND(\psi\IMPL\forall x\varphi)$\@.  Eliminating
$\rightarrow$ yields $((\NEG\forall
x\varphi)\OR\psi)\AND(\NEG\psi\OR\forall x\varphi)$\@.  To move the
quantifiers to the front, we have to push the first negation inward.
Finally, we obtain $\exists x\forall x'
((\NEG\varphi\OR\psi)\AND(\NEG\psi\OR\varphi[x'/x]))$ assuming that
$x$ does not occur free in $\psi$, and $x'$ does not occur free in
$\varphi$ and $\psi$\@.  We have not only doubled the length of the
formula but we have also doubled the number of quantifiers.  We want
to \emph{eliminate} quantifiers and have ended up doubling our work.

In analogy to the maximum of the lengths of the quantifier blocks of a
formula in prenex normal form, we define the \emph{quantifier block
  length} of the formula $\varphi$ as
\begin{equation*}
  \qbl(\varphi):=
  \max\set{\qbl_Q(\psi)}{Q\in\{\exists,\forall\}
    \text{ and }\psi\text{ is a subformula of }\varphi}
  \,, 
\end{equation*}
where
\begin{equation*}
  \qbl_Q(\varphi):=\begin{cases}
    \qbl_{\overline{Q}}(\psi) 
    & \text{if $\varphi=\neg\psi$,}
    \\
    \qbl_Q(\psi_1)+\qbl_Q(\psi_2) 
    & \text{if $\varphi=\psi_1\oplus\psi_2$ with $\oplus\in\{\wedge,\vee\}$,}
    \\
    \qbl_{Q}(\neg\psi_1\vee\psi_2) 
    & \text{if $\varphi=\psi_1\rightarrow\psi_2$,}
    \\
    \qbl_Q((\psi_1\rightarrow\psi_2)\wedge(\psi_2\rightarrow\psi_1))
    & \text{if $\varphi=\psi_1\leftrightarrow\psi_2$,}
    \\
    1+\qbl_Q(\psi) &\text{if $\varphi=Qx\psi$,}
    \\
    0 
    & \text{otherwise,}
  \end{cases}
\end{equation*}
for $Q\in\{\exists,\forall\}$\@.  

\begin{theorem}
  \label{thm:qe*_bounds}
  For every $\varphi\in\pa$ of length $n$, there is a logically
  equivalent $\psi\in\qf$ such that
  \begin{align*}
    &|\Terms(\psi)| 
    \leq
    n^{(\qbl(\varphi)+1)^{\qa(\varphi)}}
    &\qquad\text{and}\qquad
    &|\Divs(\psi)| 
    \leq
    n^{1+(\qbl(\varphi)+1)^{\qa(\varphi)+2}}
    \\
    &\maxCoef(\psi) 
    <
    a^{2^{2\qn(\varphi)}}
    &\qquad\text{and}\qquad
    &\maxDiv(\psi) 
    <
    a^{2^{2\qn(\varphi)}}
    \,,
  \end{align*}
  and
  \begin{equation*}
    \maxConst(\psi) <
    ba^{2^{3\qn(\varphi)}n^{1+(\qbl(\varphi)+1)^{\qa(\varphi)+2}}}
    \,,
  \end{equation*}
  where $a>\max\{2,\maxCoef(\varphi),\maxDiv(\varphi)\}$ and
  $b>\max\{2,\maxConst(\varphi)\}$\@.
\end{theorem}
\begin{proof}
  We require that variables are not reused in $\varphi$, \ie, the set
  of free variables of $\varphi$ is disjoint from the set of bound
  variables and the bound variables are pairwise distinct.  Note that
  this can be achieved by replacing quantified variables by fresh
  variables.  Such a variable renaming can increase the number of distinct
  atomic formulas. 
  However, the number of atomic formulas after such a renaming still
  is less than or equal to the length of the original formula.  Note
  that $n\geq\max\{2,|\Terms(\varphi)|, |\Divs(\varphi)|\}$\@.
  
  We construct the formula $\psi\in\qf$ in $\qa(\varphi)$ steps. Let
  $\varphi_0:=\varphi$\@. Let $0<k\leq\qa(\varphi)$ and assume that
  after the $(k-1)$st step we have produced the formula
  $\varphi_{k-1}$\@.  Let $\Phi$ be the set of maximal subformulas
  $\vartheta$ of $\varphi_{k-1}$ with $\qa(\vartheta)\leq 1$ and where
  variables are either only existentially quantified or universally
  quantified.
  We can assume without loss of generality that every formula in
  $\Phi$ is in prenex normal form and that
  $\Phi=\{\vartheta_1,\dots,\vartheta_m\}$\@.  For $1\leq i\leq m$,
  let $\xi_i\in\qf$ be the logically equivalent formula to
  $\vartheta_i$ from Lemma~\ref{lem:qe_bounds}\@.
  We replace in $\varphi_{k-1}$ every $\vartheta_i$ by $\xi_i$\@.
  We obtain the formula $\varphi_k$ that is logically equivalent
  to $\varphi$ and $\qa(\varphi_k)=\qa(\varphi)-k$\@.
  For $k=\qa(\varphi)$, we define $\psi:=\varphi_k$\@.

  For the formula $\varphi_k$, we have that
  \begin{align*}
    \Terms(\varphi_k)
    \subseteq
    \Terms(\varphi_{k-1})\setminus
    \Big(\bigcup_{1\leq i\leq m}\Terms_+(\vartheta_i)\Big)
    \cup
    \bigcup_{1\leq i\leq m}
    \big(\Terms(\xi_i)\setminus\Terms_-(\vartheta_i)\big)
    \,.
  \end{align*}
  Since variables are not reused in $\varphi$, it follows that
  \begin{equation*}
    |\Terms(\varphi_k)|\leq
    |\Terms(\varphi_{k-1})|-\sum_{1\leq i\leq m}|\Terms_+(\vartheta_i)|+
    \sum_{1\leq i\leq m} |\Terms_+(\vartheta_i)|^{\qn(\vartheta_i)+1}
    \,.
  \end{equation*}
  It is straightforward to show that the left hand side has its
  maximum when $m=1$ and
  $|\Terms_+(\vartheta_1)|=|\Terms(\varphi_{k-1})|$\@.  Analogously to
  the step case in the proof of Lemma~\ref{lem:qe*_bounds} for
  formulas in prenex normal form, it follows that
  $|\Terms(\varphi_k)|\leq n^{(\qbl(\varphi)+1)^{k+1}}$ under the
  assumption that $|\Terms(\varphi_{k-1})|\leq
  n^{(\qbl(\varphi)+1)^k}$\@.
  
  We can argue similarly for $|\Divs(\varphi_k)|$\@. Similar as in the
  proof of Lemma~\ref{lem:qe*_bounds} for formulas in prenex normal
  form we obtain the upper bounds for $\maxCoef(\varphi_k)$,
  $\maxDiv(\varphi_k)$, and $\maxConst(\varphi_k)$\@.
\end{proof}

\subsection{Main Result}
\label{subsec:main_result}

We now prove our main result: The upper bound on the automata size of
the minimal DWA for Presburger arithmetic formulas.
\begin{theorem}
  \label{thm:main_result}
  The size of the minimal DWA for a formula $\varphi\in\pa$ of length
  $n$ is at most $2^{n^{(\qbl(\varphi)+1)^{\qa(\varphi)+4}}}$\@.
\end{theorem}
\begin{proof}
  Since we measure the length of integers linearly, we have that the
  absolute value of every integer occurring in $\varphi$ is bounded by
  $n$\@. It holds that $n> \maxConst(\varphi)$, $n>\maxCoef(\varphi)$,
  and $n>\maxDiv(\varphi)$\@.
  
  For $\qn(\varphi)=0$, we have that the size of the minimal DWA is at
  most $2^n$\@.  For every atomic formula $\alpha_i$ of length $n_i$
  in $\varphi$, we can build a DWA of size at most $n_i$ by using the
  constructions in~\S\ref{subsec:(in)equations}
  and~\S\ref{subsec:divisibility}\@. Applying the product construct
  yields a DWA of size at most $\prod_{1\leq i\leq m} n_i\leq
  2^{\sum_{1\leq i\leq m} n_i}\leq 2^n$, where $m$ is the number of
  atomic formulas in $\varphi$\@.

  In the following, assume that $\qn(\varphi)\geq 1$ and, therefore, 
  we have that $\qa(\varphi)\geq 1$ and $\qbl(\varphi)\geq 1$\@.
  For the sake of readability, we define $a:=\qa(\varphi)$ and
  $\ell:=\qbl(\varphi)$\@.
  From Theorem~\ref{thm:qe*_bounds} it follows that there is a
  logically equivalent $\psi\in\qf$ with
  \begin{equation*}
    |\Terms(\psi)| \leq   
    n^{(\ell+1)^a}
    \qquad\text{and}\qquad
    |\Divs(\psi)| \leq 
    n^{1+(\ell+1)^{a+2}}
    \,.
  \end{equation*}
  Upper bounds on $\maxCoef(\psi)$, $\maxDiv(\psi)$, and
  $\maxConst(\psi)$ are
  \begin{equation*}
    \maxCoef(\psi), \maxDiv(\psi) <
    n^{2^{2\qn(\varphi)}}\leq
    2^{2^{2a\ell}\log_2 n}\leq
    2^{n^{1+2a\ell}}
  \end{equation*}
  and
  \begin{align*}
    \maxConst(\psi) 
    < 
    n^{1+2^{3\qn(\varphi)}n^{1+(\ell+1)^{a+2}}}
    \leq
    2^{n^{3+3a\ell+(\ell+1)^{a+2}}}
    \leq
    2^{n^{(\ell+1)^{a+1}+(\ell+1)^{a+2}}}
    \,.
  \end{align*}
  Note that $n\geq 2$, $a\ell\geq\qn(\varphi)$, and $x^{y}=2^{y\log_2
    x}$, for $x\geq 1$ and $y\geq 0$\@.
  
  Assume that there are $r\leq n$ free variables in $\varphi$\@.
  Since every term in $\psi$ contains at most the free variables of
  $\varphi$, the sum of the absolute values of the coefficients in a
  term is bounded by $n\cdot n^{2^{2\qn(\varphi)}}\leq
  2^{n^{2+2a\ell}}<2^{n^{3+3a\ell}}$\@.  With
  Theorem~\ref{thm:bound_qf_formula} at hand, we know that the size of
  the minimal DWA for $\psi$ is at most
  \begin{align*}
    \Big(2+2\cdot 2^{n^{(\ell+1)^{a+1}+(\ell+1)^{a+2}}}\Big)^{|\Terms(\psi)|}\cdot 
    \maxDiv(\psi)^{|\Divs(\psi)|}
    \,.
  \end{align*}
  From
  \begin{align*}
    \Big(2+2\cdot 2^{n^{(\ell+1)^{a+1}+(\ell+1)^{a+2}}}\Big)^{|\Terms(\psi)|} 
    \leq
    2^{n^{(\ell+1)^{a+1}+(\ell+1)^{a+2}+(\ell+1)^a}}
    \leq
    2^{n^{(\ell+1)^{a+3}}}
  \end{align*}
  and
  \begin{align*}
    \maxDiv(\psi)^{|\Divs(\psi)|}
    \leq 2^{n^{2+2a\ell+(\ell+1)^{a+2}}}
    \leq
    2^{n^{2(\ell+1)^a+(\ell+1)^{a+2}}}
    \leq
    2^{n^{(\ell+1)^{a+3}}}
  \end{align*}
  we conclude that the size of the minimal DWA for $\varphi$ is at
  most $2^{n^{(\ell+1)^{a+4}}}$\@.
\end{proof}

Theorem~\ref{thm:main_result} does not change if we measure the length
of integers logarithmically and not linearly.  The only change is that
the maximal absolute integer in $\varphi$ is now smaller than $2^n$\@.
We have to adjust the bounds on $\maxCoef(\psi)$, $\maxDiv(\psi)$, and
$\maxConst(\psi)$\@. For instance, we still have that
\begin{equation*}
  \maxCoef(\psi)<
  (2^n)^{2^{2\qn(\varphi)}}=2^{n2^{2\qn(\varphi)}}\leq
  2^{n^{1+2\qa(\varphi)\qbl(\varphi)}}
  \,.
\end{equation*}
 We argue analogously for
$\maxDiv(\psi)$ and $\maxConst(\psi)$\@.

\begin{corollary}
  \label{cor:aut_size}
  Let $\pa_c$ be the set of PA formulas with at most $c\geq 0$ quantifiers.
  The size of the minimal DWA for each $\varphi\in\pa_c$ is at most
  $2^{n^{\BigOh(1)}}$, where $n$ is the length of $\varphi$\@.
\end{corollary}
\begin{proof}
  If $\qn(\varphi)\leq c$ then $\qa(\varphi)\leq c$ and
  $\qbl(\varphi)\leq c$\@. Since $c$ is fixed the claim follows
  directly from Theorem~\ref{thm:main_result}\@.
\end{proof}

We want to remark that Theorem~\ref{thm:main_result} and
Corollary~\ref{cor:aut_size} only give upper bounds on the sizes of
the minimal DWAs for PA formulas.  If the Boolean connectives and the
quantifiers are handled by standard automata constructions, like
complementation and subset construction, and the DWAs are minimized
after every automata construction step, it may be the case that the
whole construction uses one exponent more space.  The reason is that
an exponential blow-up can occur each time the subset construction is
applied.  It is an open question whether the standard automata
constructions already suffice to construct a DWA in
$2^{n^{(\qbl(\varphi)+1)^{\qa(\varphi)+4}}}$ space or time, for a
given $\varphi\in\pa$ of length $n$\@.  It is also open if there are
more efficient automata constructions than the standard ones for
constructing DWAs for PA formulas.

\section{A Worst Case Example}
\label{WORST}

We give a worst case example that shows that our upper bound on the
automata size is tight.
We use the formulas $\Prod_n(x,y,z)$ defined by Fischer and
Rabin~\citeyear{Fischer_Rabin.1974}, for $n\geq 0$\@.  It holds that
\begin{equation*}
  \Zrepresents{\Prod_n}=
  \set{(a,b,c)\in\Nat}{ab=c\text{ and }
    a,b,c<\!
    \prod_{\substack{p\text{ is prime and}\\p<f(n+2)}}\!\!p}
  \,,
\end{equation*}
where $f(n):=2^{2^n}$\@.  Note that it follows from the Prime Number
Theorem that
\begin{equation*}
  \prod_{\substack{p\text{ is prime and}\\p<f(n+2)}}\!\!p\geq
  2^{f(n)^2}=
  2^{f(n+1)}
  \,.
\end{equation*}
Fischer and Rabin looked at the structure $(\Nat,+)$ and not at
$\frakZ$, but it is straightforward to adapt the definition of
$\Prod_n(x,y,z)$ to $\frakZ$\@.
For $n\geq 0$, the length of $Prod_n$ and the number of quantifier
alternations is linear in $n$\@.  The quantifier block length is
constant, \ie, there is a $c\geq 0$ such that for all $n\geq 0$,
$\qbl(\Prod_n)=c$\@. By Theorem~\ref{thm:main_result} we know that the
minimal DWA for $\Prod_n$ has at most $2^{2^{2^{\BigOh(n)}}}$ states.

Before we prove the lower bound on the automata size for the formulas
$\Prod_n$, we need the following lemma.
\begin{lemma}
  \label{lem:mult_prefix}
  Let $\ell\geq 1$\@. For all $z\in\Nat$ with $\varrho^{\ell-1}\leq
  z\leq\varrho^{\ell}-2$, there are $x,y,z'\in[\varrho^{\ell}]$
  such that $xy = \varrho^{\ell} z+z'$\@.
\end{lemma}
\begin{proof}
  Assume that $\varrho^{\ell-1}\leq z\leq\varrho^{\ell}-2$\@.
  Let $x,y\in[\varrho^{\ell}]$ with $xy\geq \varrho^{\ell} z$ and
  $xy-\varrho^{\ell}z$ is minimal. Note that it is always possible to
  find $x,y\in[\varrho^{\ell}]$ with $xy\geq \varrho^{\ell}z$ since
  for $x=y=\varrho^{\ell}-1$, we have that
  \begin{equation*}
    xy 
    =
    (\varrho^{\ell}-1)^2 
    =
    \varrho^{2\ell}-2\varrho^{\ell}+1 
    \geq
    \varrho^{\ell}(\varrho^{\ell}-2)
    \geq 
    \varrho^{\ell}z
    \,.
  \end{equation*}
  
  Let $z':=xy-\varrho^{\ell}z$\@.  We have to show that
  $z'\in[\varrho^{\ell}]$\@.  Since $xy\geq \varrho^{\ell}z$ we have
  that $z'\geq 0$\@.  For the sake of absurdity, assume that $z'\geq
  \varrho^{\ell}$\@.  It follows that
  \begin{equation*}
    (x-1)y=xy-y = \varrho^{\ell}z+z'-y
    \geq
    \varrho^{\ell}z
  \end{equation*}
  since $y<\varrho^{\ell}$ and $z'\geq\varrho^{\ell}$\@.  This
  contradicts the minimality of $xy-\varrho^{\ell}z$ since
  $xy>(x-1)y\geq \varrho^{\ell}z$\@.
\end{proof}

Our proof for the lower bound on the automata size for a formula
$\Prod_n$ is based on the following lemma about the set
\begin{equation*}
  \MULT_m:=\set{(a,b,c)\in\Int^3}{a,b\in[\varrho^m]\text{ and }ab=c}
  \,, 
\end{equation*}
for $m\geq 0$\@.
\begin{lemma}
  \label{lem:mbit_multiplication}
  Let $m\geq 0$\@.  Every DWA representing $\MULT_m$ has at least
  $\varrho^m$ states.
\end{lemma}
\begin{proof}
  For $m=0$, the claim is trivial. In the following, assume that $m>0$
  and that $\autA=(Q,\Sigma^3,\delta,q_{\rmI},F)$ is a DWA
  representing $\MULT_m$. Let $K$ be the set of words of the form
  $(0,0,0)(0,0,b_{m-1})\dots(0,0,b_0)\in(\Sigma^3)^*$ with
  $b_{m-1}\not=0$ and $b_0\leq\varrho-2$\@.  Let $w\in K$ and let $z$
  be the integer that is encoded by the third track of $w$\@.  It
  holds that
  \begin{equation*}
    \varrho^{m-1}\leq z\leq\varrho^m-2
    \,.
  \end{equation*}
  From Lemma~\ref{lem:mult_prefix} it follows that there are
  $x,y,z'\in[\varrho^m]$ such that
  \begin{equation*}
    xy=\varrho^mz+z'
    \,.
  \end{equation*}
  We conclude that for every prefix $u$ of a word in $K$ there is a
  word $v\in(\Sigma^3)^*$ such that $\toints{uv}\in\MULT_m$\@.
  
  Now, let $L$ be the set of all prefixes of $K$\@. 
  Let $u,u'\in L\setminus\{\lambda\}$ with $u\not=u'$\@. Moreover, let
  $v\in(\Sigma^3)^*$ with $\toints{uv}\in\MULT_m$\@.
  The first and second tracks of $uv$ and $u'v$ encode both the pair
  $(x,y)$\@. The third tracks of $uv$ and $w'v$ are different. It
  follows that $\toints{u'v}\notin\MULT_m$ and hence,
  $\widehat{\delta}(q_{\rmI},u)\not=\widehat{\delta}(q_{\rmI},u')$\@.
  We conclude that the DWA $\autA$ must have a distinct state for
  every word in $L$\@.
  
  In the following, we determine the cardinality of $L$\@.  For
  $0\leq i\leq m+1$, let $L_i:=\set{w\in L}{\length{w}=i}$\@.  We have
  that $L_0=\{\lambda\}$, $L_1=\{(0,0,0)\}$,
  $L_2=\set{(0,0,0)b}{b\in\Sigma\setminus\{0\}}$, $L_i=\set{wb}{w\in
    L_{i-1}\text{ and }b\in\Sigma}$, for $3\leq i\leq m$, and
  $L_{m+1}=K$\@.  It holds that
  \begin{equation*}
    \begin{array}{@{}r@{\,}c@{\,}l@{}}
      |L|&=&
      |L_0|+|L_1|+|L_2|+|L_3|+\dots+|L_m|+|L_{m+1}| 
      \\
      &=&
      1 + 1 + (\varrho-1) + (
      \varrho-1)\varrho + \dots + (\varrho-1)\varrho^{m-2} + 
      (\varrho-1)\varrho^{m-1}-2
      \\
      &=&
      \varrho^m
      \,.
      \qed
    \end{array}
  \end{equation*}
\end{proof}

\begin{theorem}
  \label{thm:worst_case}
  Let $n\geq 0$\@. The size of every DWA representing
  $\Zrepresents{\Prod_n}$ is at most least
  $2^{\bigl\lfloor\frac{f(n+1)}{2\log_2\varrho}\bigr\rfloor}$\@.
\end{theorem}
\begin{proof}
  Assume that for $n\geq 0$, there is a DWA $\autB$ with less than
  $2^{\bigl\lfloor\frac{f(n+1)}{2\log_2\varrho}\bigr\rfloor}$ states
  representing the set $\Zrepresents{\Prod_n}$\@.  Let
  $m:=\big\lfloor\frac{f(n+1)}{2\log_2\varrho}\big\rfloor$\@.  It holds
  that $\MULT_m\subseteq \Zrepresents{\Prod_n}$ since
  $(\varrho^m-1)^2<\varrho^{2m}=2^{2m\log_2\varrho}\leq2^{f(n+1)}$\@.
  It is straightforward to construct from $\autB$ a DWA
  representing the set $\MULT_m$ that has as many states as $\autB$ by
  making some of the accepting states in $\autB$ non-accepting. This
  contradicts Lemma~\ref{lem:mbit_multiplication}\@.  
\end{proof}

\begin{remark}
  We make the following remarks on nondeterministic word automata and
  alternating word
  automata~\cite{Brzozowski_Leiss.1980,Chandra_Kozen_Stockmeyer.1981}\@.
  \begin{enumerate}[(i)]
  \item The proof of Theorem~\ref{thm:worst_case} carries over to
    nondeterministic word automata.  That means, that we obtain the
    same lower bound for nondeterministic word automata as for DWAs
    although nondeterministic word automata can sometimes be
    exponentially more succinct than DWAs.  
  \item A lower bound for the number of states of alternating word
    automata for the formula $\Prod_n$ is at least
    $\bigl\lfloor\frac{f(n+1)}{2\log_2\varrho}\bigr\rfloor$\@.  This
    lower bound follows by contradiction from the remark~(i) above and
    the fact that an alternating word automaton can be translated to
    an equivalent nondeterministic word automaton with exponentially
    more states.  \ignore{ Note that alternating word automata can be
      double exponentially more succinct than DWAs~\cite{???}\@.
      However, for $\Prod_n$ they can be at most exponentially more
      succinct.
    
    In the following, we sketch the construction of an alternating
    word automaton for $\Prod_n$ that has exponentially less states
    than any DWA for $\Prod_n$\@.  \dots}
  \end{enumerate}
\end{remark}

\section{Conclusion}
\label{CONCL}

We analyzed the automata-theoretic approach for deciding Presburger
arithmetic and established a tight upper bound on the automata size.
Moreover, we improved the automata constructions
in~\cite{Boigelot.1999,Wolper_Boigelot.2000,Ganesh_Berezin_Dill.2002}
for equations and inequations and proved that our automata
constructions are optimal.

The main technique to prove the upper bound on the automata size was
to relate deterministic word automata with the formulas constructed by
a quantifier elimination method.
This technique can also be used to prove upper bounds on the sizes of
minimal automata for other logics that admit quantifier elimination
and where the structures are automata
representable~\cite{Khoussainov_Nerode.1994,Blumensath_Graedel.2000,Rubin.2004},
\ie, these structures are provided with automata for deciding equality
on the domain and the atomic relations of the structure.  Prominent
examples are the mixed first-order theory over the structure
$(\mathbb{R},\Int,\mathbin{<},+)$~\cite{Boigelot_Jodogne_Wolper.2001,Weispfenning.1999}
and the first-order theory of
queues~\cite{Rybina_Voronkov.2001,Rybina_Voronkov.2003}\@.

\bibliographystyle{acmtrans}
\bibliography{prsb}

\begin{received}
  Received Month Year; revised Month Year; accepted Month Year
\end{received}

\end{document}